\documentclass{aa}  

\usepackage{graphicx}
\usepackage{txfonts}
\begin{document}

   \title{A 600\,kpc complex radio source at the center of Abell 3718 discovered by the EMU and POSSUM surveys}

   \author{F. Loi\inst{1},
          M. Brienza\inst{2,3,4},
          C.~J. Riseley\inst{3,4,5},
          L. Rudnick\inst{6},
          W. Boschin\inst{7,8,9},
          L. Lovisari\inst{2,10},
          E. Carretti\inst{4},
          B. Koribalski\inst{5,11},
          C. Stuardi\inst{3,4},
          S. P. O'Sullivan\inst{12},
          A. Bonafede\inst{3,4},
          M. D. Filipovi\'c\inst{11},
          A. Hopkins\inst{13}.}
    \authorrunning{F. Loi et al.}
    \titlerunning{A 600\,kpc complex radio source at the center of A3718 discovered by EMU and POSSUM surveys}
   \institute{
        INAF--Osservatorio Astronomico di Cagliari, via della Scienza 3, Selargius, Italy\\
        \email{francesca.loi@inaf.it}
        \and 
        INAF -- OAS Osservatorio di Astrofisica e Scienza delle Spazio di Bologna, Via Gobetti 93/3, 40129, Bologna, Italia
        \and 
        Dipartimento di Fisica e Astronomia, Università degli Studi di Bologna, via P. Gobetti 93/2, 40129 Bologna, Italy
        \and 
        INAF – Istituto di Radioastronomia, via P. Gobetti 101, 40129 Bologna, Italy
        \and 
        Australia Telescope National Facility (ATNF), CSIRO Astronomy and Space Science, P.O. Box 76, Epping, NSW 1710, Australia 
        \and 
        Minnesota Institute for Astrophysics, University of Minnesota, 116 Church St. SE, Minneapolis, MN 55455, USA
        \and 
        Fundaci\'on G. Galilei - INAF (Telescopio Nazionale Galileo), Rambla J. A. Fern\'andez P\'erez 7, E-38712 Bre\~na Baja (La Palma), Spain
        \and 
        Instituto de Astrof\'{\i}sica de Canarias, C/V\'{\i}a L\'actea s/n, E-38205 La Laguna (Tenerife), Spain
        \and 
        Departamento de Astrof\'{\i}sica, Univ. de La Laguna, Av. del Astrof\'{\i}sico Francisco S\'anchez s/n, E-38205 La Laguna (Tenerife), Spain
        \and 
        Center for Astrophysics $|$ Harvard $\&$ Smithsonian, 60 Garden Street, Cambridge, MA 02138, USA
        \and 
        School of Science, Western Sydney University, Locked Bag 1797, Penrith, NSW 2751, Australia
        \and 
        School of Physical Sciences and Centre for Astrophysics \& Relativity, Dublin City University, Glasnevin D09 W6Y4, Ireland
        \and 
        Australian Astronomical Optics, Macquarie University, 105 Delhi Rd, North Ryde, NSW 2113, Australia
        }

   \date{Received December 7, 2022; accepted January **, ****}

 
  \abstract
   {Multifrequency studies of galaxy clusters are crucial for inferring their dynamical states and physics. Moreover, these studies allow us to investigate cluster-embedded sources, whose evolution is affected by the physical and dynamical condition of the cluster itself. So far, these kinds of studies have been preferentially conducted on clusters visible from the northern hemisphere due to the high-fidelity imaging capabilities of ground-based radio interferometers located there.}
   {In this paper, we conducted a multifrequency study of the poorly known galaxy cluster Abell 3718. We investigated the unknown origin of an extended radio source with a length of $\sim612$ kpc at 943 MHz detected in images from the Evolutionary Map of the Universe (EMU) and POlarisation Sky Survey of the Universe's Magnetism (POSSUM) surveys.}
   {We analyzed optical and X-ray data to infer the dynamical state of the cluster and, in particular, the merger activity. We conducted a radio spectral index study from 943\,MHz up to 9\,GHz. We also evaluated the polarization properties of the brightest cluster-embedded sources to understand if they are related to the radio emission observed on larger scales.}
   {The cluster appears to be in a relaxed dynamical state, but there is clear asymmetry of the X-ray surface brightness distribution perpendicular to the direction of the largest angular extension of the radio source. The morphology of the cluster radio emission observed from 900 MHz to 9 GHz shows a system composed of a northern compact radio source and a southern radio galaxy whose jets are bent in the direction of an ultra-steep ($\alpha\approx$3.6), thin (few tens of kpc) arc of radio emission between the first two radio sources. The spectral index gradient along the radio source and the polarization images at high frequency suggest that the thin arc is an extension of the southern radio galaxy, which may have been energized by interacting with the X-ray gas. An additional structure extending to the northwest from the southern radio galaxy may be an unusual truncated radio jet that either failed to expand or faded away due to energy losses.  Deeper X-ray and radio observations are needed to better constrain the physics at play in this cluster.}
   {}


   \maketitle
%
\section{Introduction}
Galaxy clusters host an incredible realm of radio source diversity: star forming galaxies, radio galaxies, remnant radio galaxies, radio phoenixes, radio halos, radio relics, and mini-halos \citep[see e.g.,][for the most recent review of the field]{vanweeren2019}. With the advent of the new generation of radio telescopes paving the way for the upcoming Square Kilometre Array (SKA), we are detecting ever more peculiar sources that challenge our historic taxonomy and provide new opportunities for understanding the complex physics at work in clusters \citep[see e.g.,][]{govoni2019,Botteon2020b,Botteon2020c,Biava2021,rudnick2021,Riseley2022b}. Indeed, the detection and the classification of cluster-embedded radio sources can provide crucial information about the evolution of both clusters and cluster galaxies and how the sources interact with the environment. \
\begin{figure*}
    \centering
    \includegraphics[width=0.49\textwidth]{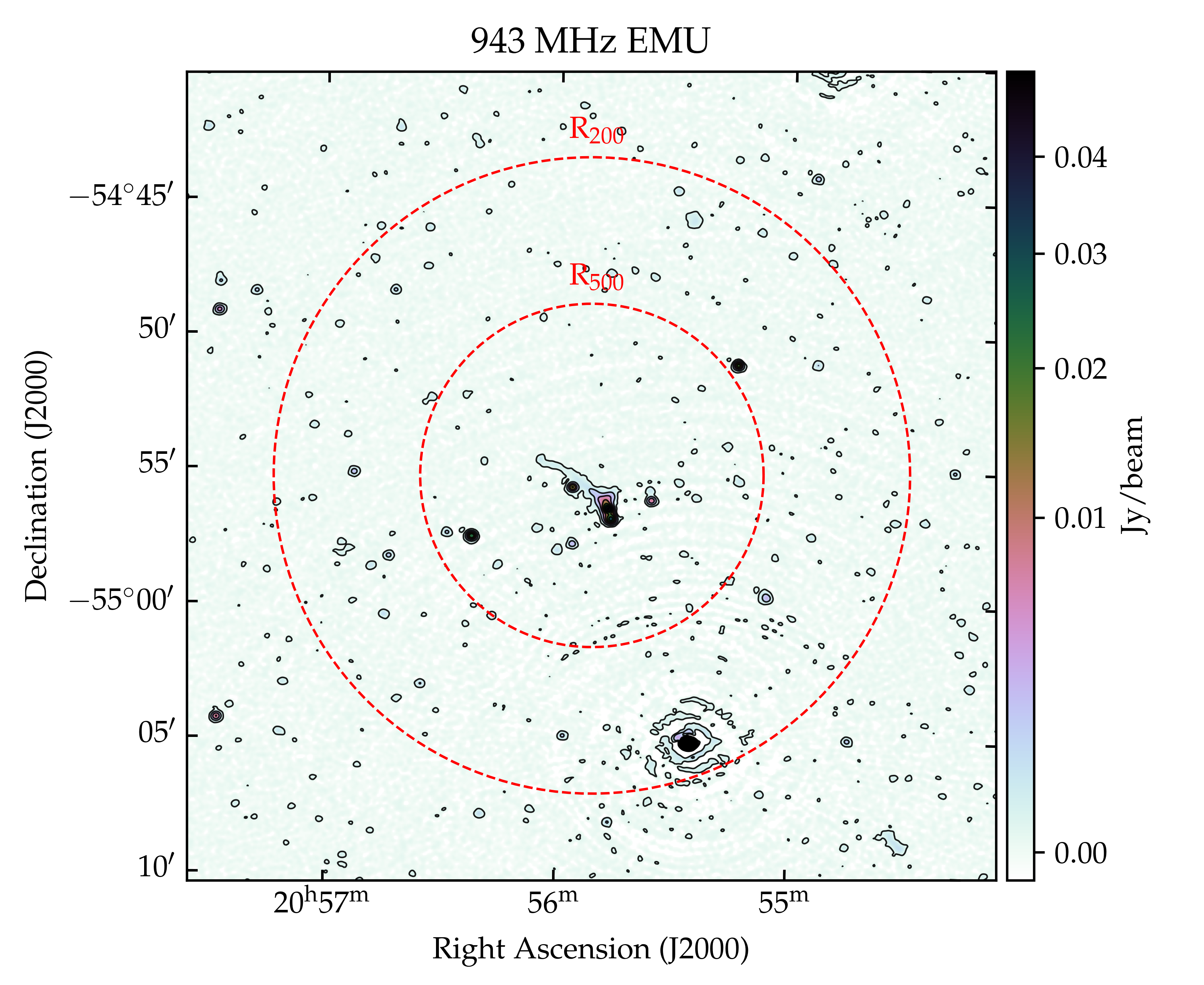}
    \includegraphics[width=0.49\textwidth]{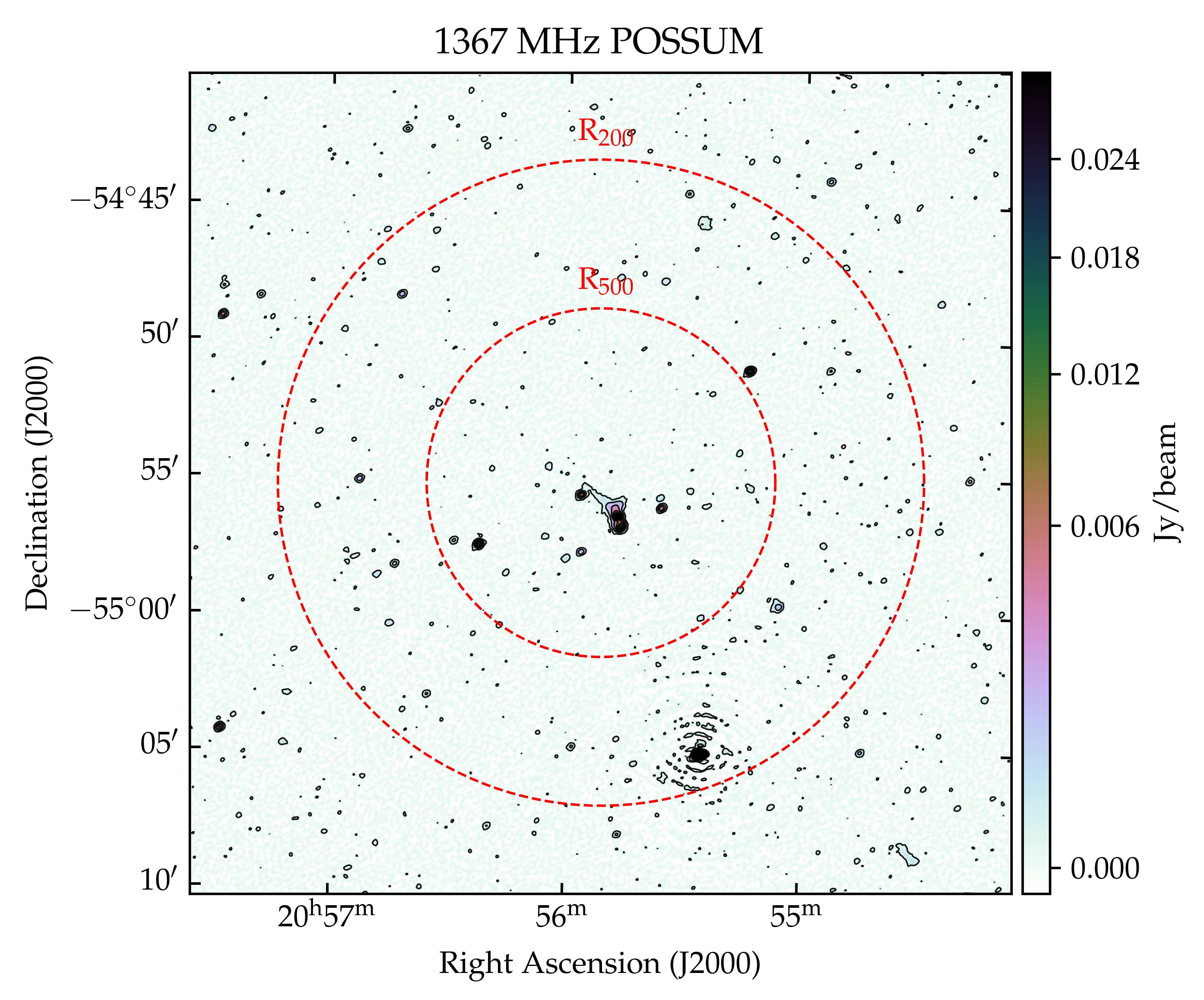}
    \caption{ASKAP images of the galaxy cluster A3718. Left: 943\,MHz EMU total intensity image. The beam size is 13.96$\times$10.89\,arcsec$^2$ and BPA=$-$58.15\,deg. Right: 1.367\,MHz POSSUM total intensity image. The beam size is 10.00$\times$7.56\,arcsec$^2$ and BPA=$-$54.63 deg. In both images contours start at 3$\sigma$ and increase by a factor of $\sqrt{2}$ with $\sigma$ equal to 40\,$\muup$Jy/beam and 30\,$\muup$Jy/beam in the EMU and POSSUM image, respectively. The colorbar is in logaritmic scale. The two dashed circles represent the cluster R$_{200}$ and R$_{500}$ radii equal to 1.8\,Mpc and 974\,kpc, respectively.}
    \label{fig:emu}
\end{figure*}

For instance, ram-pressure stripping phenomena on cluster galaxies are a well known example of how the intracluster medium (ICM) can dramatically change a galaxy's morphology and therefore its evolution, forming the so-called jellyfish galaxies \citep[e.g.][]{Smith2010,Fumagalli2014, Ebeling2014}.
Similarly, the interaction between active radio jets and the ICM can bend the radio lobes of a radio galaxy, shaping the well-known head-tail radio galaxies \citep[see e.g.,][]{Ryle1968}. Radio phoenixes form instead when the Active Galactic Nucleus (AGN) activity ceases and the non-thermal component is compressed \citep{ensslin2001,ensslin2002}.

Merging events in clusters can generate shock waves and turbulence that (re-)accelerate particles and both amplify and spread preexisting magnetic fields \citep{markevitch2007}. This non-thermal component in the ICM can give rise to diffuse synchrotron sources not associated with optical galaxies, namely, radio halos and relics. Radio halos are Mpc-scale emission located at the cluster center, while radio relics are usually found in the cluster outskirts. In some cases, however, radio relics have been observed close to the cluster center. An example is the ``chair'' relic in MACS J0717.5+3745 \citep[see e.g.,][]{Bonafede2018,rajpurohit2022}. 
Some radio relics show a spatial connection with the lobes of a radio galaxy \citep{Bonafede2014,Shimwell2015,vanWeeren2017NatAs,Stuardi2019,HyeongHan2020}, suggesting that radio relics may be powered by a re-accelerated electron population rather than an electron population being accelerated directly from the thermal pool.

Multiwavelength radio observations are crucial to classifying radio sources. Indeed, they allow us to better understand the structure of these sources as well as what is triggering the observed emission (either merging phenomena, compression, AGN, etc.) and therefore obtain important hints surrounding a cluster's physics. For example, radio relics usually show a spectral index gradient perpendicular to the shock front (i.e., the relic elongation if the relic is seen edge-on), while head-tail radio galaxies have a flat spectral index at the core location and steeper values along the lobes as particles age. 
Optical and X-ray data can be used to complement the view on the cluster and cluster source physics by revealing the presence of substructures. 
Moreover, since radio relics are typically triggered by shock waves, sharp discontinuities in the X-ray images (in particular in the surface brightness, temperature, density, and pressure) can help identify these kinds of sources. 
Conversely, cold fronts correspond to a discontinuous drop in temperature across the interface paired with a jump in density but not in pressure (as in shocks), and they are often observed bounding minihalos \citep[see e.g.,][]{Mazzotta2001ApJ...555..205M,Mazzotta2001astro.ph..8476M,Mazzotta2008,Ghizzardi2010,Rossetti2013}. However, emerging evidence from deep observations with the SKA Pathfinders and Precursors has complicated our understanding, with some minihalos extending far beyond the boundaries of known cold fronts \citep[e.g.,][]{Biava2021,Riseley2022a}.

In this paper, we report the results of a detailed multifrequency analysis of the galaxy cluster Abell 3718 (hereafter, A3718; R.A. 20h55m51.5s, Dec. $-$54d55m12s, z=0.139). 
The cluster has a mass of $\rm M_{500}=3.01\times 10^{14} \, M_{\odot}$ and a radius of ${\rm R_{500}=974\,kpc}$ \citep{piffaretti2011}, that is, the radial distance from the cluster center at which the cluster density is 500 times the critical density of the Universe. This cluster has remained poorly studied in the literature. Thus, we present and analyze the first available radio observations of this cluster. Our new observations allow us to detect peculiar extended radio emission in this cluster for the first time, covering a projected distance of $\sim$612\,kpc. 

The A3718 cluster was observed with the Australian SKA Pathfinder \citep[ASKAP;][]{mcconnell2016,hotan} at 943\,MHz as part of the Evolutionary Map of the Universe \citep[EMU;][]{norris2011} survey. In particular, observations took place during the EMU Phase I Pilot Survey \citep{norris2021} performed in ASKAP Band~1 and centered on a frequency of 947\,MHz. The full EMU survey will conduct a census of the radio continuum source population across the (predominantly) southern sky and is predicted to detect up to about 70 million radio sources \citep{norris2011}. The left panel of Figure \ref{fig:emu} shows the 943\,MHz EMU image in colors and contours. 
The same sky area was observed by ASKAP in Band 2 at a central frequency of 1367\,MHz during the Phase I Pilot Survey of the POlarisation Sky Survey of the Universe's Magnetism \citep[POSSUM;][]{gaensler2010}. The corresponding image is shown on the right of Figure \ref{fig:emu}. The goal of POSSUM is to investigate the magnetic fields in the Milky Way as well as in galaxies, in clusters, and in the overall intergalactic medium. Both surveys detected the extended radio emission, although the full extent was not recovered in the POSSUM image due to its lower signal-to-noise ratio.

\begin{figure*}
    \centering
    \includegraphics[width=0.8\textwidth]{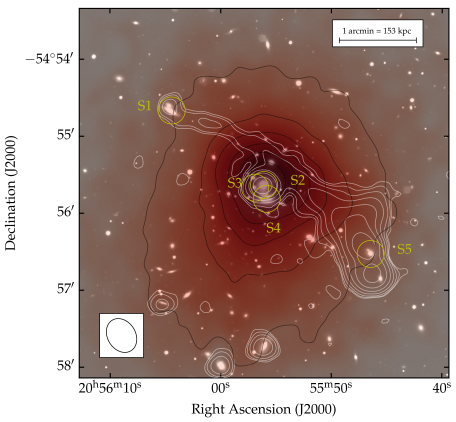}
    \caption{Multiwavelength image of galaxy cluster A3718. The background color map shows our composite $i-$, $r-,$ and $g-$band image from the DES. Red colors and black contours trace the X-ray surface brightness measured in the 0.3--7\,keV range by \emph{XMM-Newton}, and white contours show the 943\,MHz EMU total intensity surface brightness starting from 3$\sigma$ with $\sigma$=40\,$\muup$Jy/beam and increasing by a factor of $\sqrt{2}$. See the text for more details.}
    \label{fig:multi}
\end{figure*}

With its large field of view, high resolution, and good sensitivity to extended, low-surface brightness emission, ASKAP is ideally suited to performing detailed studies of galaxy clusters and, indeed, has already had significant impact in the field \citep[e.g.,][]{HyeongHan2020,Wilber2020,anderson2021,brueggen2021,reiprich2021,dimascolo2021,Duchesne2021b,Duchesne2021c,Duchesne2021a,veronica2022,venturi2022,loi2022,Riseley2022b}. Such studies have revealed new insights into diffuse radio sources in galaxy clusters that have furthered --- and in many cases strongly challenged --- our understanding of the physics of the ICM.

In this work, we aim to determine the nature of the peculiar extended radio emission and, in particular, of the thin radio arc. The morphology of the radio arc suggests two possible interpretations: the radio source is either a radio relic caused by a spherical shock wave or a tail associated with a radio galaxy possibly shaped by interactions with the ICM.

The paper is organized as follows. Optical and X-ray data and images are presented in Section 2. In Section 3, we show the radio data and images from EMU and POSSUM as well as follow-up observations performed with the Australia Telescope Compact Array \citep[ATCA;][]{FraterBrooks1992} at 5.5 GHz and 9 GHz. Our X-ray analysis is presented in Section 4. In Section 5, we analyze the spectral properties of the radio sources. In Section 6, we discuss the results, and in Section 7 we report our conclusions. Throughout the paper we assume a flat $\rm\Lambda$CDM cosmology with $H_0=67.4$\,km/s/Mpc, $\rm\Omega_M=0.315$ and $\rm\Omega_{\Lambda}=0.685$ \citep{planck2020}. At the cluster redshift (i.e., $z=0.139$), an angular scale of 1\,arcsec corresponds to a physical distance of 2.54332\,kpc.

\section{Optical and X-ray data and images}
Information about the dynamical state of a cluster can be inferred from the analysis of the cluster galaxies and of the X-ray emission associated with the ICM. In what follows we present the optical and the X-ray data and images used to constrain the cluster's dynamical state.

\subsection{Optical data and images}
We searched the literature for galaxies belonging to A3718. We
found five galaxies whose spectroscopic redshifts suggest that they are fiducial cluster members. These galaxies are identified in Figure \ref{fig:multi}, which shows a superposition of the Dark Energy Survey (DES) image, the X-ray emission (see next section), and the EMU image. We found two additional likely cluster-member galaxies in the DES \citep[]{des2021} based on the corresponding photometric redshifts. 

Since we had only five cluster-member galaxies with measured spectroscopic redshifts, we instead used the DES photometric redshifts to search for the presence of substructures in A3718. We constructed our catalog by selecting the galaxies in the A3718 field that have photometric redshifts in the range $0.08<z_{\rm phot}<0.18$ and uncertainties $\Delta$z$_{\rm phot}\leq$0.05. We then applied the Voronoi tessellation and percolation (VTP) procedure to this catalog \citep[e.g.,][]{ramella2001,barrena2005}.
The VTP is a nonparametric technique that is sensitive to galaxy structures of any shape and symmetry. 
The VTP detected A3718 as a unique, statistically significant galaxy overdensity (at the 99.9\% c.l.; 46 galaxies assigned to the cluster) with no hint of substructures in the plane of the sky. 
However, we cannot exclude the possibility that the cluster has a substructure along the line of sight, as we did not have a large sample of spectroscopically confirmed cluster-member galaxies with which to explore this possibility.

\subsection{X-ray data and images}
At X-ray wavelengths, A3718 was observed with \emph{XMM-Newton} with a total exposure time of 43 ks (ObsID:0675010901). The observation was performed in full-frame mode for the Metal Oxide Semi-conductor (MOS) cameras and extended full-frame mode for the pn detector, all using the thin filter. Although the target of the observation was the galaxy cluster SPT-CL J2056$-$5459 (a cluster at redshift $z=0.72$) and A3718 is located 9 arcmin off axis, the data are sufficient for a good characterization of the X-ray emission associated with A3718.  

We  retrieved the observation data files and reprocessed them with the \emph{XMM-Newton} Science Analysis System (SAS) v19.0.0. We used tasks "emchain" and "epchain" to generate calibrated event files from the raw data. Throughout the analysis, we only used events with PATTERN$<$13 for MOS data and with PATTERN$<$5 for pn data. In addition, for all cameras, we excluded all events next to Charged Coupled Device (CCD) edges and bad pixels (i.e., FLAG==0).

The data sets were then cleaned for periods of high background due to soft protons following the two-stage filtering process described in \cite{Lovisari2011}. To briefly summarize, we first extracted a light curve with bins of 100\,s bins in the 10–12 (12–14) keV energy band for MOS (pn) and  fitted a Poisson distribution to the histogram of this light curve. We then excluded all the time intervals deviating by more than 2$\sigma$ from the mean of the distribution. The new event lists were then refiltered in a second pass as a safety check for possible flares with soft spectra (e.g., \citealt{DeLuca2004}). In this case, each light curve was made with 10\,s bins in the full 0.3--10 keV band.
The remaining exposure times after cleaning were $\sim$40 ks for MOS and $\sim$30 ks for pn. Point-like sources were detected using the task "edetect-chain" and checked by eye before excluding them from the analysis.  
The background event files were cleaned by applying the same PATTERN selection, flare rejection, and point-source removal used for the observation events.

The X-ray image presented in this work was obtained in the 0.3--7 keV band using data from both the MOS and pn detectors and smoothed to a resolution of 6\,arcsec. From the raw image, we subtracted all the components of the background (i.e., particle and cosmic) and then applied the vignetting correction by dividing with the exposure map.

\begin{table*}
\centering
\caption{Summary of radio observations.}
\label{table:freq}
\begin{tabular}{c c c c c c}
   \hline
   central freq. [GHz] & BW [GHz] & Instrument & $uv-$min & rms & res. [arcsec$^2$]\\
   \hline
   0.943 & 0.288 & ASKAP-EMU & 69$\lambda$ & 40\,$\muup$Jy/beam & 13.96$\times$10.89\\
   1.367 & 0.144 & ASKAP-POSSUM & 100$\lambda$ & 30\,$\muup$Jy/beam & 10$\times$7.56\\
   5.5 & 2 & ATCA & 1118$\lambda$ & 7\,$\muup$Jy/beam & 1.656$\times$3.364\\
   9.0 & 2 & ATCA & 1830$\lambda$ & 6\,$\muup$Jy/beam & 1.069$\times$2.255\\
   \hline
\end{tabular}
\end{table*}

In Figure \ref{fig:multi} we present a color composite image of A3718 from our multiwavelength data. We show an optical DES image with our {\it XMM-Newton} 0.3--7\,keV map  overlaid with our 943\,MHz EMU map. From Figure \ref{fig:multi}, the cluster appears to be elongated along a NW--SE axis, with a flattening of the X-ray emission along the NW edge visible in the tighter contour spacing, potentially indicating motion of the thermal gas. We analyze this further in Section 4.

\section{Radio data and images}
In this section, we present the data and images obtained with ASKAP at 943 and 1367 MHz and ATCA at 5.5 and 9 GHz. All of the data are summarized in Table \ref{table:freq}.

\subsection{ASKAP data and images}
The EMU and POSSUM observations used in this work have the following ASKAP Scheduling Block IDs (SBIDs): SBID~9351 and SBID~9975, respectively. Fully (direction-independent) calibrated mosaic images were downloaded through the Commonwealth Scientific and Industrial Research Organisation (CSIRO) ASKAP Science Data Archive \citep[CASDA;][]{Chapman2017,Huynh2020} portal\footnote{\url{https://research.csiro.au/casda/}}.  
The EMU\footnote{EMU data: Norris, Ray; Filipovic, Miroslav; Huynh, Minh; Hotan, Aidan; Rudnick, Lawrence; Manojlovic, Perica; Gurkanuygun, Gulay; Parkinson, David; Macgregor, (2019): ASKAP Data Products for Project AS101 (ASKAP Pilot Survey for EMU): images and visibilities. v1. CSIRO. Data Collection. http://hdl.handle.net/102.100.100/164553?index=1} 
data cover the frequency range between 800\, MHz and 1088\,MHz, while the POSSUM\footnote{POSSUM data: Gaensler, Bryan; Mcclure-Griffiths, Naomi; Purcell, Cormac; Huynh, Minh; Hotan, Aidan; Rudnick, Lawrence; Anderson, Craig; Kaczmarek, Jane; Heald, George; West, Jennifer (2020): ASKAP Data Products for Project AS103 (ASKAP Pilot Survey for POSSUM): images and visibilities. v1. CSIRO. Data Collection. http://hdl.handle.net/102.100.100/326209?index=1} 
data are between 1295\,MHz and 1439\,MHz. Data reduction was performed using the \textsc{ASKAPSoft} pipeline \citep{askapsoft}.\\

As can be seen in Figures \ref{fig:emu} and \ref{fig:multi}, the long radio emission appears to connect (in projection) the northern cluster member S1 with the southern cluster galaxy S5 and beyond, covering a projected distance of $\sim$612\,kpc. The two bright peaks in the radio surface brightness close to the galaxy S5 (R.A. $313.94358\deg$, Dec. $-54.94225\deg$) could be two radio lobes associated with the galaxy. This radio galaxy covers a distance of $\sim$200\,kpc. The length of the faint arc is $\sim$370\,kpc. Its width decreases going from SW to NE, from $\sim$13\,arcsec to 8\,arcsec, and it reaches a minimum of $\sim$6.5\,arcsec. These scales correspond to 33, 21, and 16\,kpc, respectively, and are below the beam size. It is therefore possible that we are detecting only the tip of the arc emission, which could indeed be wider than what is observed in the EMU image. The faint arc is not completely retained at higher frequency in the POSSUM image.

The extended radio emission is therefore composed of three main structures. The first is a compact radio galaxy associated with S1. The second structure is a southern radio galaxy associated with S5, which we will henceforth refer to as "S5RG", and the third is a faint radio arc connecting the two structures in projection. The radio arc is $\sim$30\,kpc from the X-ray peak. An additional structure was clearly detected in both images extending from the radio galaxy S5RG toward the northwest of A3718. We suggest that this structure could be a radio jet that failed to expand or faded away due to energy losses. Another possibility is that the radio jet bends toward the Earth, making it undetectable.

\begin{figure*}
\centering
    \includegraphics[width=0.45\textwidth]{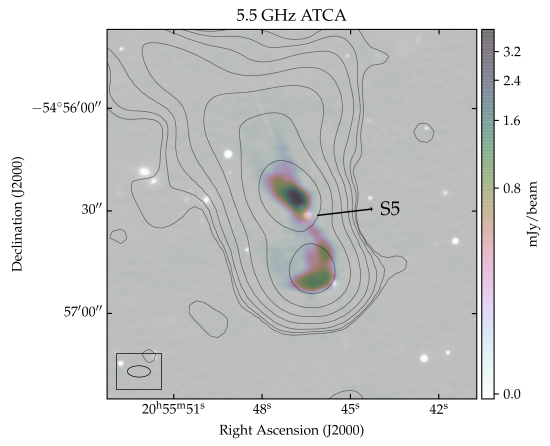}
    \includegraphics[width=0.45\textwidth]{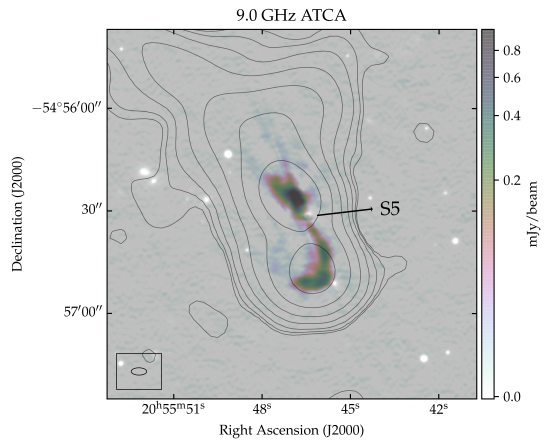}
    \caption{ATCA total intensity images of the S5RG. Left: 5.5 GHz image with a beam resolution of 1.656$\times$3.364\,arcsec$^2$ and BPA=89.2\,deg. Right: 9.0\,GHz image with a beam resolution of 1.069$\times$2.255\,arcsec$^2$ and BPA=89.6\,deg. In both the images we show the 943\,MHz EMU image in gray contours, the DES image in gray scale colors, and the beam sizes in the bottom left-corner.}
    \label{fig:atca}
\end{figure*}

\subsection{ATCA data and images}
Higher-resolution observations were required to attempt to determine whether there is any association between the extended radio arc detected at 943\,MHz and the southern radio galaxy S5RG. We carried out this investigation using the ATCA, with observations centered on S5, as the steep spectrum observed between 943\,MHz and 1367\,MHz (see Section 5) implied that the surface brightness would be extremely faint and challenging to recover at ATCA frequencies. 

\subsubsection{Data reduction}
We observed S5RG during semester 2021OCT (Code C3440, P.I. F.~Loi) using the 4\,cm band receiver system (4.9--10.9\,GHz) with three complementary array configurations (6A, 1.5C, and H168) to ensure we achieved high resolution while maintaining sensitivity to angular scales corresponding to the structures measured from the EMU and POSSUM data.
We used the standard reference frequencies of 5.5 and 9\,GHz for our 4\,cm band observations; at each reference frequency, the Compact Array Broadband Backend \citep[CABB;][]{Wilson2011} provides a bandwidth of 2048\,MHz at a standard 1\,MHz spectral resolution. The total observing time was of the order of 30\,hours evenly divided between the three configurations. Each observing run included scans of the primary bandpass and flux density calibrator PKS~B1934$-$638 and of the phase calibrator J2040$-$5735.

All data reduction and imaging was performed with the \textsc{miriad} software package \citep{Sault1995}, with calibration performed independently on each of the two intermediate frequency bands  using standard techniques. The radio frequency interference environment of the ATCA 4\,cm band is remarkably clean, and the flagging fraction was around 18\%.

We imaged the total intensity data using a 6000 pixel grid with a pixel size of 0.4 arcsec. The final resolution of the 5.5\,GHz and 9.0\,GHz maps are 1.7\,arcsec$\times$3.4\,arcsec with bpa=89\,deg and 1.1\,arcsec$\times$2.3\,arcsec with bpa=89\,deg, respectively. After initial imaging, we performed four rounds of phase-only self-calibration using solution intervals of five minutes for the first round, two minutes in the second round, and one minute for each of the last two rounds of self-calibration. 

The Q and U Stokes data were imaged with a 6000 pixel grid and a pixel size of 0.4 arcsec. From these data, we derived the polarized intensity and angle images. We also derived maps of the Faraday rotation measure (RM) by fitting the polarization angle against $\lambda^2$ using the \textsc{miriad} function \textsc{imrm}.

\subsubsection{Total intensity images}
Figure \ref{fig:atca} presents color composite images with the DES image in gray scale, our ATCA maps at 5.5\,GHz and 9.0\,GHz overlaid in color, and the 943\,MHz EMU surface brightness.
The two ATCA images have a beam resolution of 1.656$\times$3.364\,arcsec$^2$ and 1.069$\times$2.255\,arcsec$^2$ with BPA=89.2\,deg and 89.6\,deg, respectively, and a noise $\sigma$=7\,$\muup$Jy/beam and $\sigma$=6\,$\muup$Jy/beam, respectively.

Our ATCA data support the interpretation that the two surface brightness peaks seen in the ASKAP EMU and POSSUM maps are in fact the lobes of a radio galaxy associated with the cluster-member galaxy S5, which we previously referred to as S5RG. Our ATCA maps clearly reveal the structure of this radio galaxy, which is composed of a pair of radio lobes. Additionally, a compact radio counterpart to the optical galaxy is visible in our 9\,GHz map due to the higher resolution; at 5.5\,GHz, this compact counterpart blends into the northern lobe.

The northern lobe appears to extend further to the north in the direction of the extended tail and arc-like emission. This supports the interpretation that the faint radio arc seen at 943\,MHz is connected to the radio galaxy S5RG. The southern lobe extends toward the southwest before bending to the east and again to the northeast, also following the direction of the arc.

One further intriguing feature revealed by our high-resolution ATCA data is that the northern lobe appears to show a forked tail toward its northernmost extremity. One fork is aligned with the northeast, in the direction of the extended arc, whereas the other fork appears to extend north-northwest, in the direction of emission seen by ASKAP at lower frequencies. Magnetized filamentary structures in the ICM have been detected in a number of cases \citep[see for example][]{Rudnick2022ApJ...935..168R,Giacintucci2022ApJ...934...49G,Condon2021ApJ...917...18C,Fanaroff2021MNRAS.505.6003F}. The underlying magnetic field structure is modified, and the magnetic field line is stretched by physical mechanisms not yet completely understood \citep{Yusef-Zadeh2022arXiv221004913Y}.

\begin{figure*}
    \centering
    \includegraphics[width=0.45\textwidth]{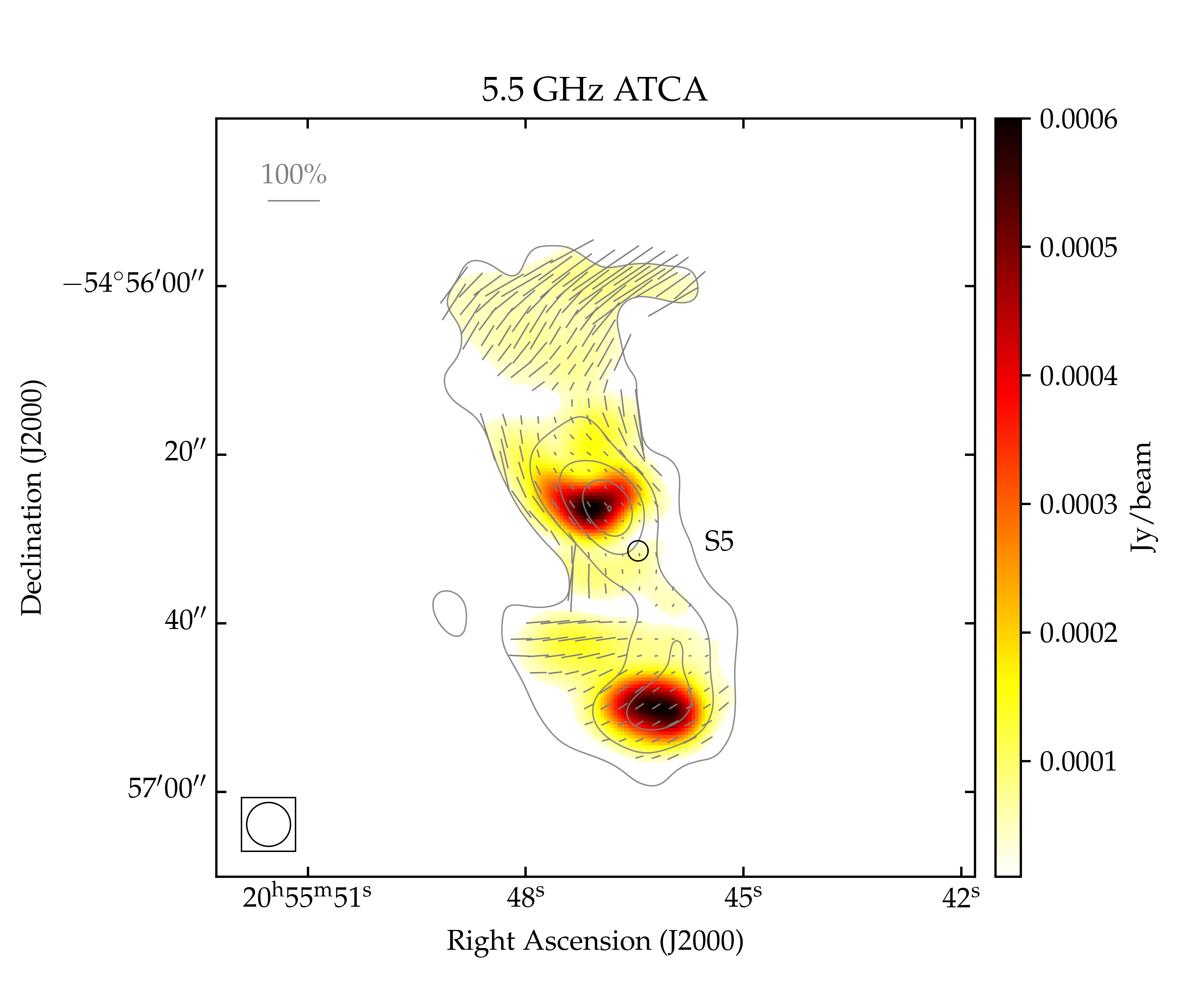}
    \includegraphics[width=0.45\textwidth]{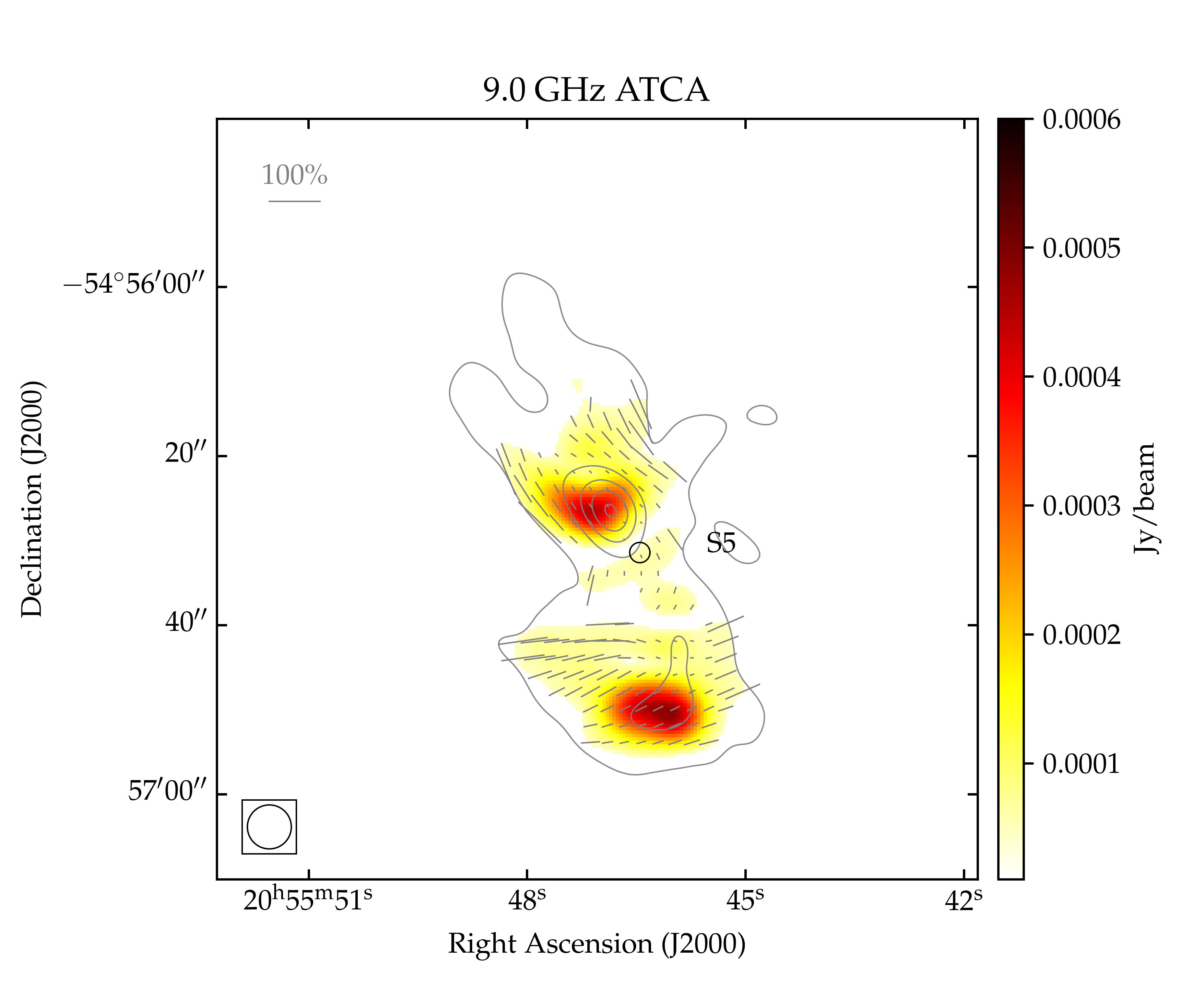}
    \caption{ATCA linearly-polarization images of the S5RG. Left: 5.5 GHz image. Right: 9.0 GHz image. In both panels we use segments to show the intensity and the orientation of the B-field vectors. A segment with fractional polarization equal to 100\% is shown in the top-left corner of the images for reference. We also overlaid the total intensity contours starting from 42\,$\muup$Jy/beam and increasing by a factor of $\sqrt{2}$. All of the images have a beam resolution of 5\,arcsec, shown in the bottom-left corner.}
    \label{fig:pol}
\end{figure*}
\begin{figure}
    \centering
    \includegraphics[width=0.45\textwidth]{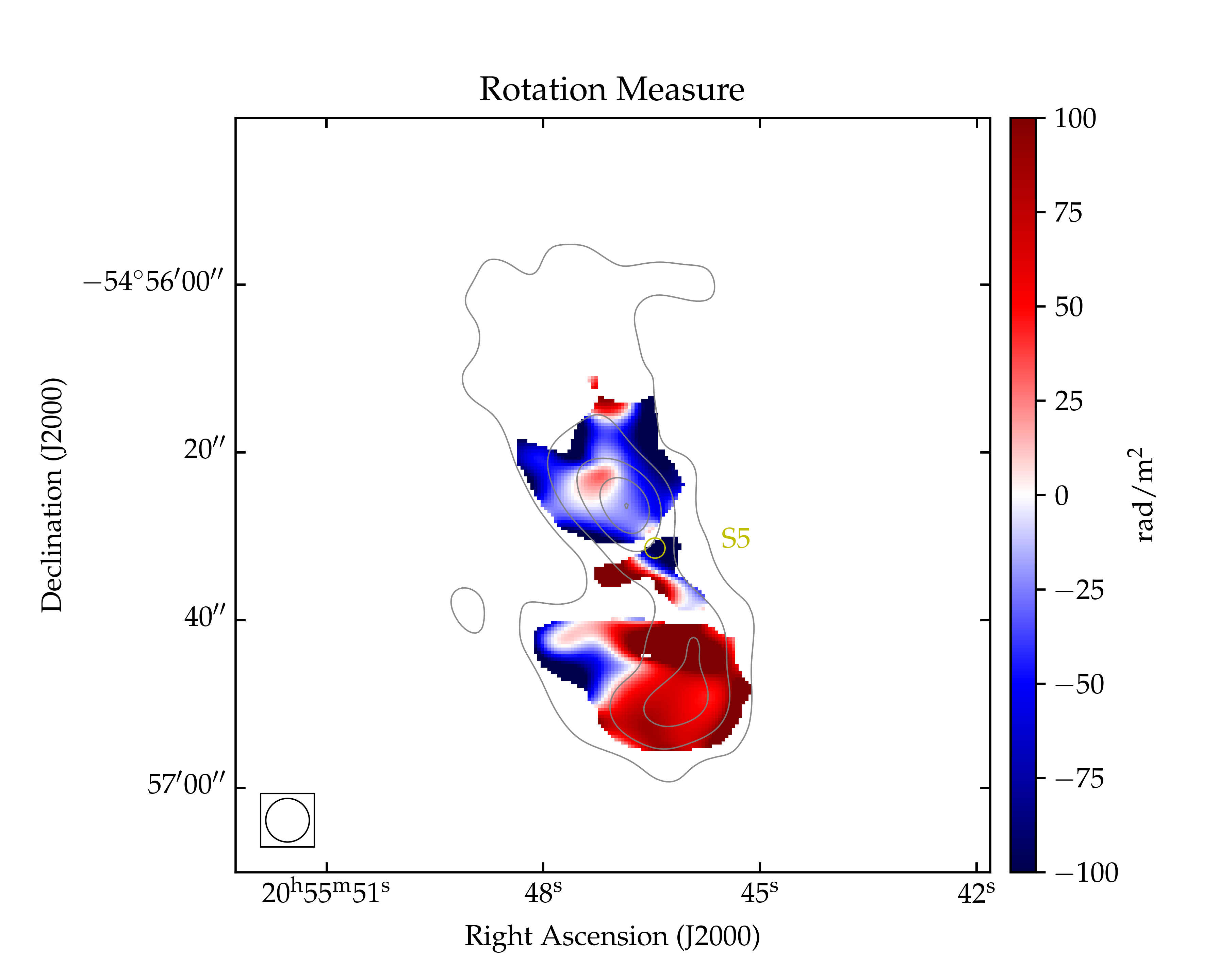}
    \caption{Rotation measure image computed from  5.5 GHz and 9.0 GHz ATCA data.}
    \label{fig:rm}
\end{figure}

\subsubsection{Polarization and rotation measure images}
We made images of the linear polarization at 5.5 and 9.0\,GHz convolving the Stokes Q and U images to a circular beam of 5\,arcsec. Figure \ref{fig:pol} shows the linearly polarized emission with polarization vectors overlaid and rotated by 90 degrees to indicate the direction of the magnetic field and the length of the vectors proportional to the fractional polarization. A vector polarized at 100\% is shown in the top-left of the figure for reference. We also included the total intensity contours. To make these images, we considered the signal with a S/N$>$4 in polarized intensity and a S/N$>$3 in total intensity. The median and standard deviation of the fractional polarization at 9.0\,GHz are $\sim$21\% and $\sim$23\%, respectively.

At 5.5\,GHz and 9.0\,GHz, the vectors should provide a very close indication of the intrinsic magnetic field orientation since Faraday rotation is negligible as $\lambda^2 \ll 1$. 
Indeed, the linearly polarized emission from cluster-embedded and background radio sources undergoes Faraday rotation from the magneto-ionic medium that is the ICM. The polarization angle rotates as a function of wavelength according to $\rm\Delta \Psi = RM \times \lambda^2$, where $\Delta\Psi$ is the observed rotation and RM is the rotation measure. The RM is defined as the path integral of the magnetic field times the thermal particle density along the path between the emitted signal and the telescope. In the ATCA images, the vectors should provide a close indication of the intrinsic magnetic field orientation, even without correction for the local RM. A value of ${\rm |RM|} = 100$\,rad/m$^2$, which is higher than we observed, would cause rotations of $\sim$6 and $\sim$17\,degrees at 9.0\,GHz and 5.5\,GHz, respectively.

At both 5.5\,GHz and 9.0\,GHz, the magnetic field vectors of the northern lobe of the radio galaxy are aligned with the direction of the associated lobe or jet. Further to the north, where the tail forks, the eastern part of the fork (which aligns with the larger-scale arc) keeps the same magnetic field orientation, whereas the western part of the fork exhibits a transverse magnetic field orientation. This could suggest helical structure within the magnetic field. At the southern lobe location, an abrupt change in the magnetic field orientation can be observed and after that an alignment along the east-west direction.

Figure \ref{fig:rm} shows the map of the RM obtained from our ATCA images at 5.5\,GHz and 9.0\,GHz. We observed negative RM values on the northern lobe and positive RMs on the southern lobe, with weighted averages of  ($-20\pm14$)\,rad/m$^2$ and ($+63\pm10$)\,rad/m$^2$,  respectively. The errors represent the weighted errors in the averages, and we did not detect any significant scatter within each lobe above that expected from the noise. The Galactic RM at this location is $(+36\pm8$)\,rad/m$^2$ \citep{Hutschenreuter2022A&A...657A..43H}\footnote{Using the CIRADA RM cutout server at \href{http://cutouts.cirada.ca/rmcutout/}{http://cutouts.cirada.ca/rmcutout/}}. The average RMs of the two lobes differ significantly from this Galactic RM value, indicating likely contributions from the ICM.

\section{X-ray analysis}
In this section, we present a detailed analysis of our \emph{XMM-Newton} data to investigate whether the properties of the radio arc are being influenced by the ICM.
In particular, the X-ray asymmetry will be better investigated.
\begin{figure*}
    \centering
    \includegraphics[width=0.9\textwidth]{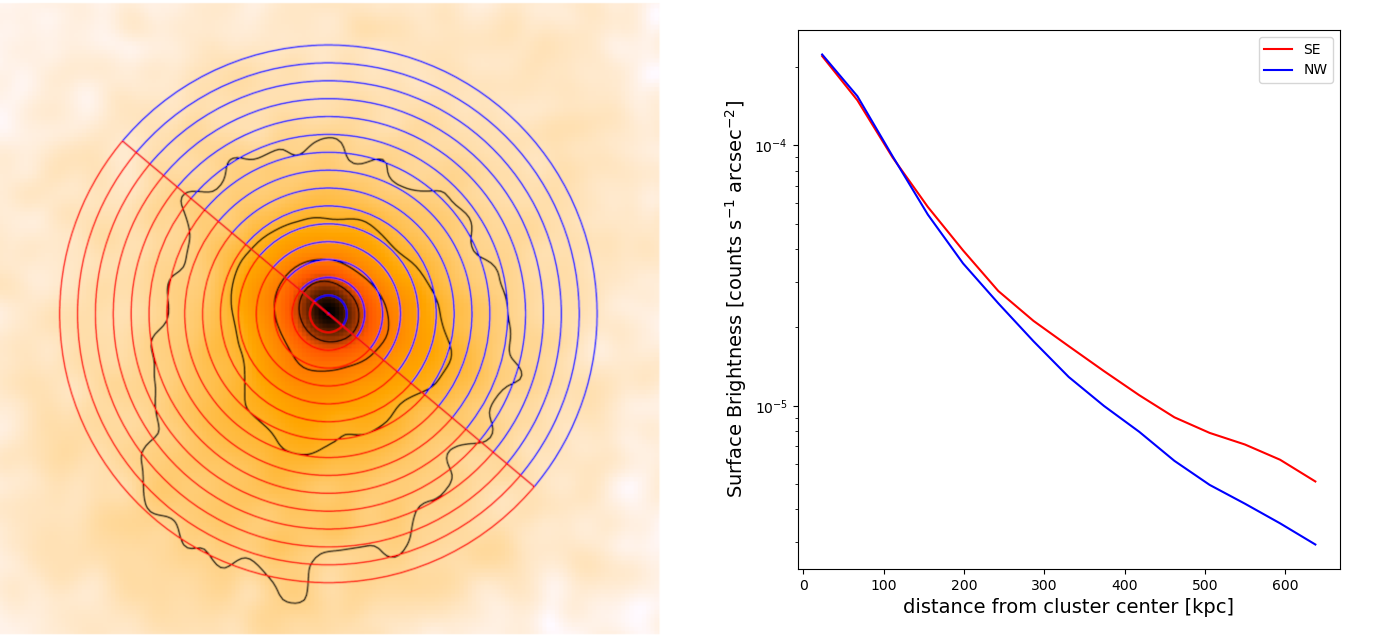}
    \caption{X-ray surface brightness image and profile. Left: X-ray surface brightness image between 0.2-7\,keV in colors and with contours. Green and red annuli start from the brightest peak of emission and extend toward the NW and SE, respectively. The inclination of the sectors is perpendicular to the radio arc. Right: X-ray surface brightness profiles corresponding to  NW and SE annuli in same color code.}
    \label{fig:panda}
\end{figure*}
\begin{figure*}
    \centering
    \includegraphics[width=0.49\textwidth]{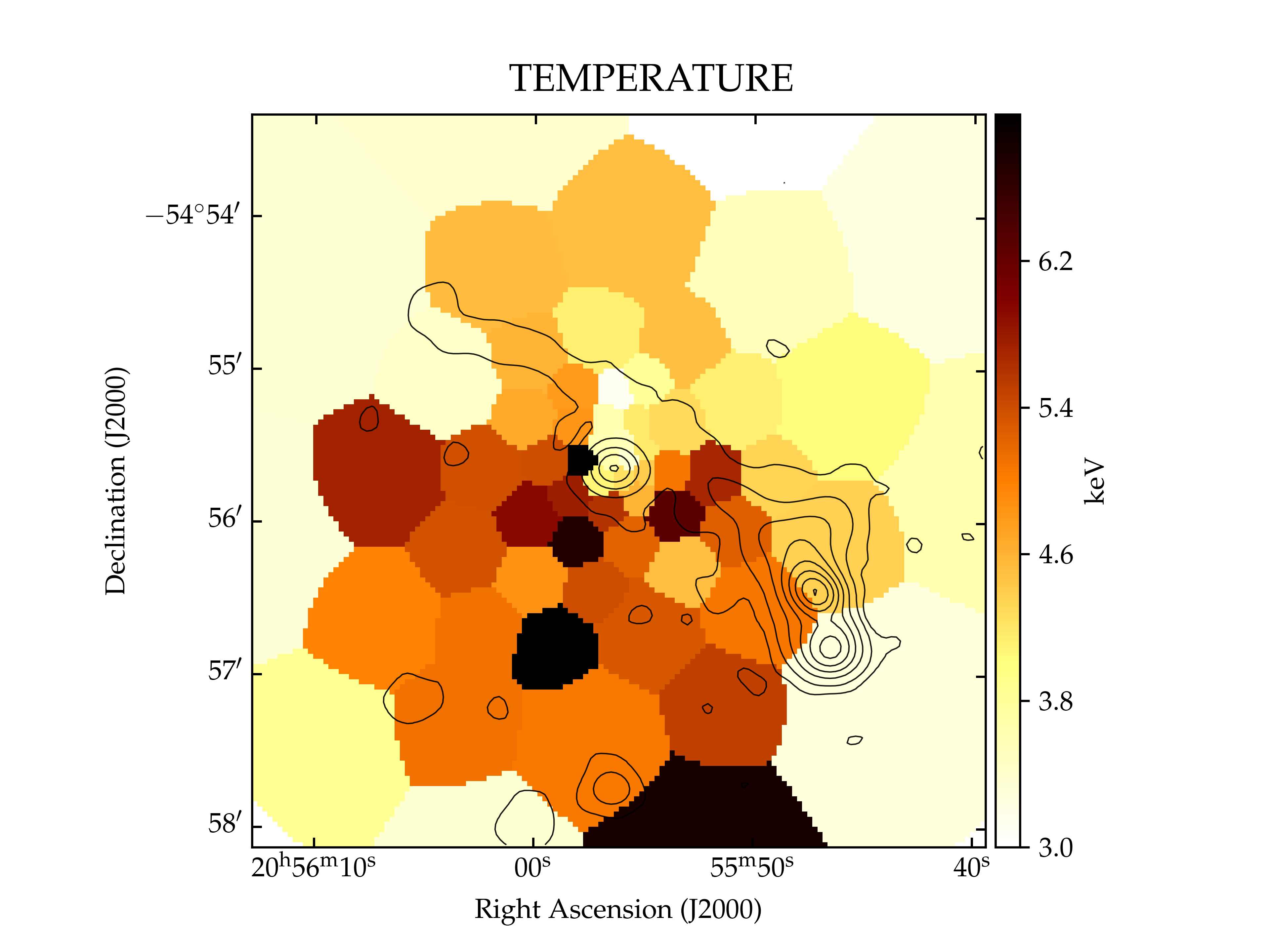}
    \includegraphics[width=0.49\textwidth]{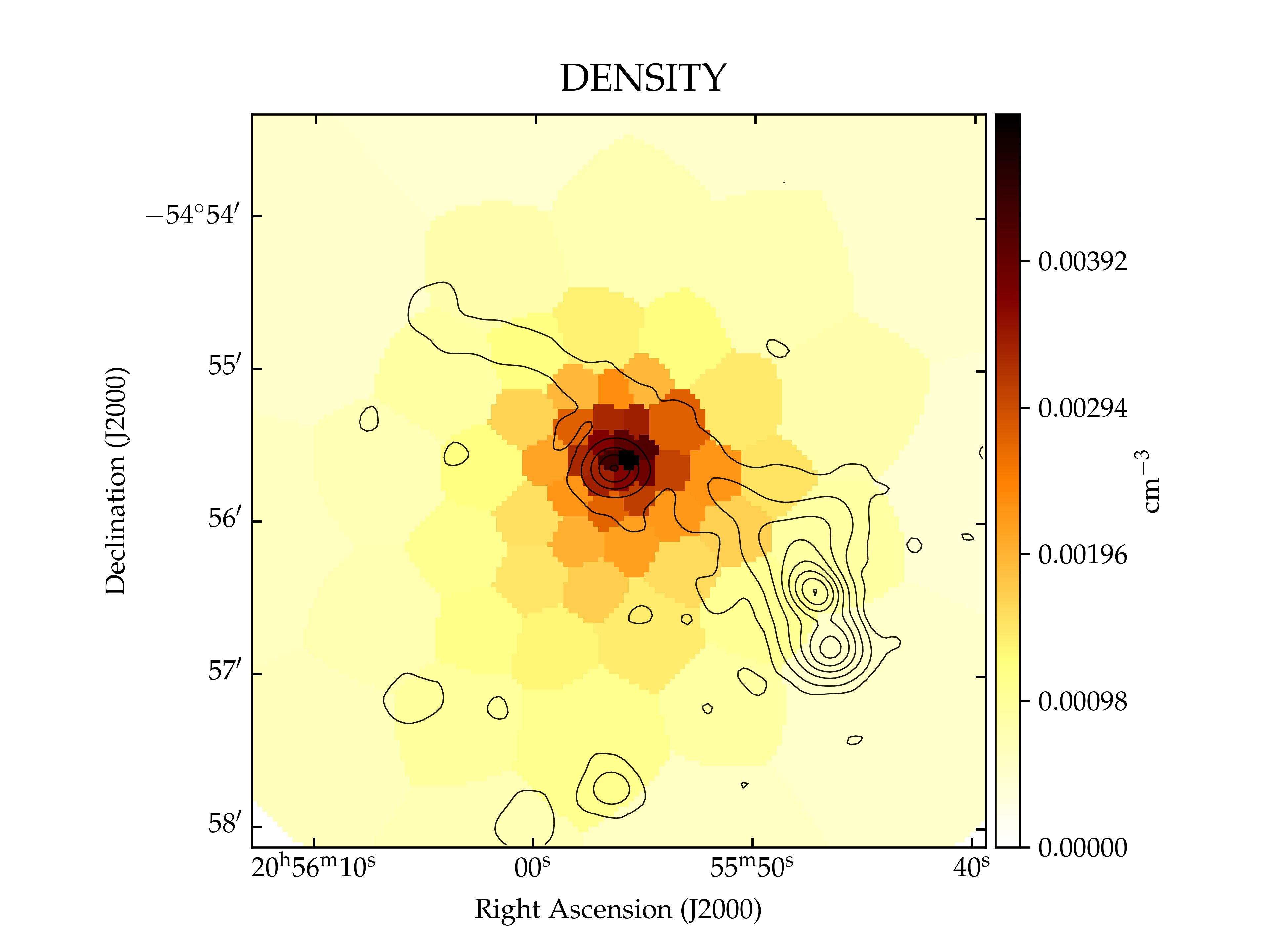}
    \includegraphics[width=0.49\textwidth]{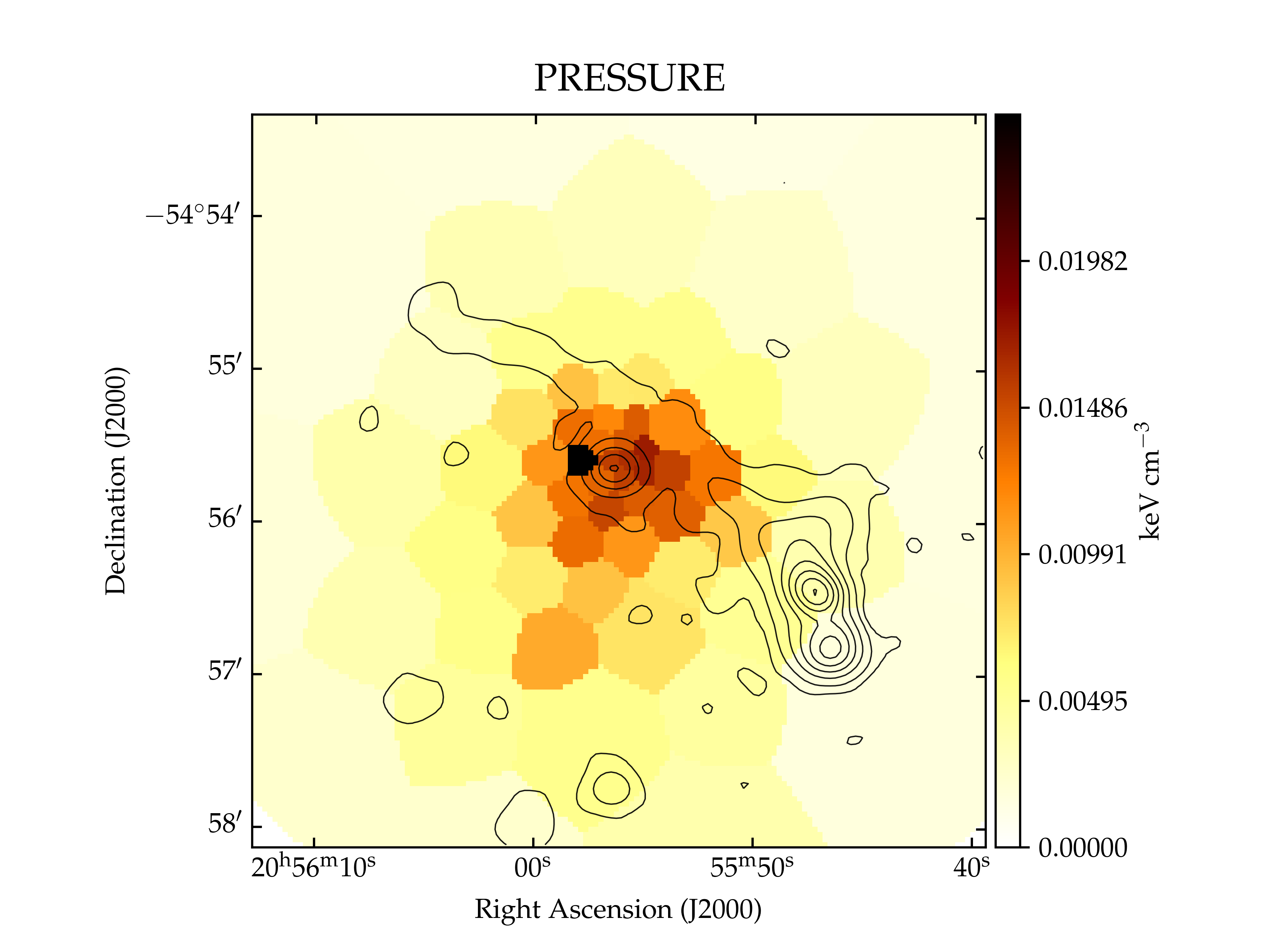}
    \includegraphics[width=0.49\textwidth]{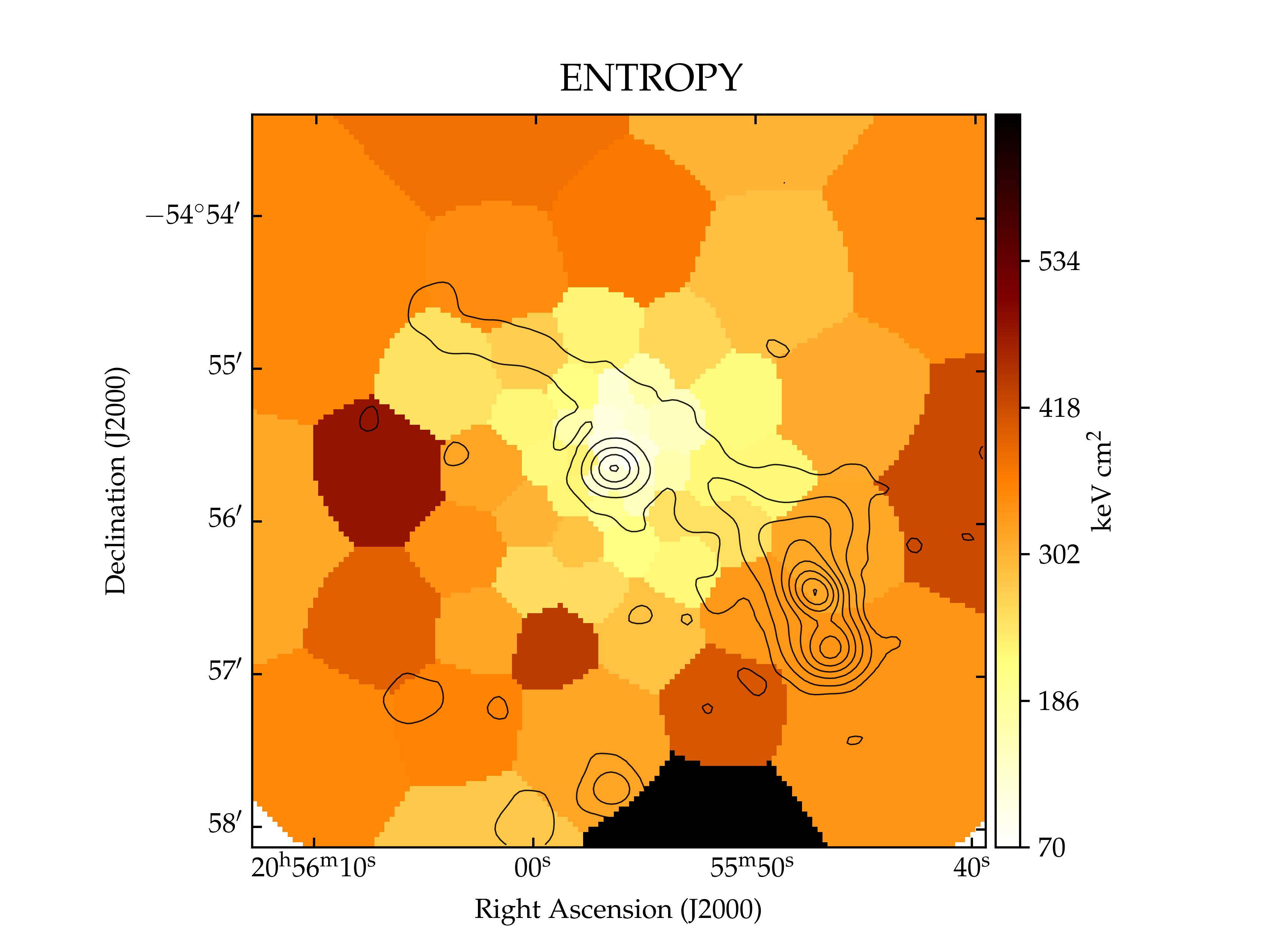}
    \caption{Projected temperature (top-left), density (top-right), pressure (bottom-left), and entropy (bottom-right) with 943\,MHz EMU contours.}
    \label{fig:voronoi}
\end{figure*}

\begin{figure*}
    \centering
    \includegraphics[width=0.49\textwidth]{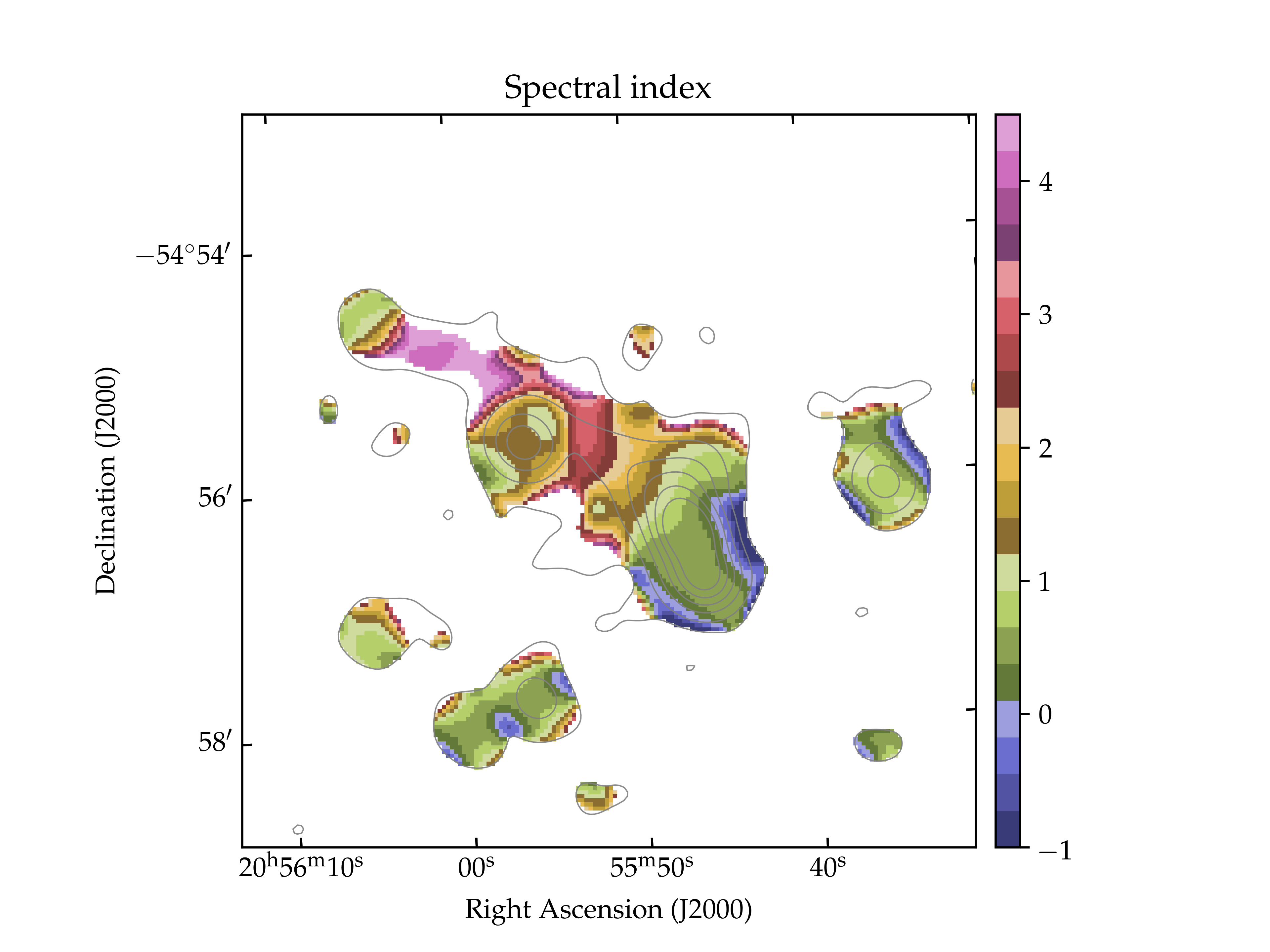}
    \includegraphics[width=0.49\textwidth]{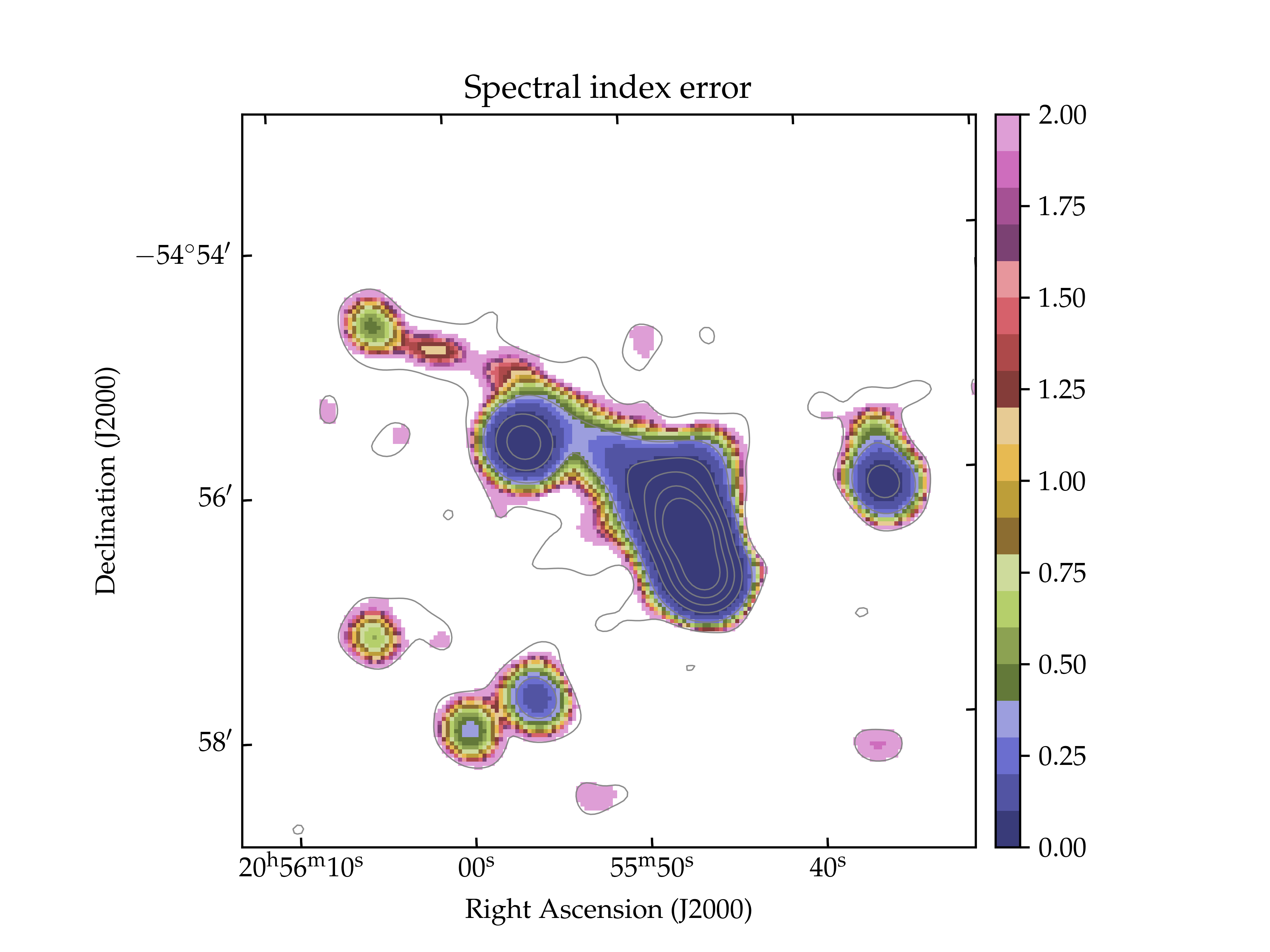}
    \caption{ASKAP spectral index map and uncertainty map computed between 943 MHz and 1.367 GHz at 13\,arcsec of resolution.}
    \label{fig:askap_spix}
\end{figure*}

\subsection{X-ray asymmetry}
The X-ray asymmetry observed in Figure \ref{fig:multi} is better shown in Figure \ref{fig:panda}. On the right of Fig. \ref{fig:panda}, we show the soft X-ray surface brightness (in the 0.7 to 2\,keV range) profiles toward the NW and SE starting from the X-ray peak. The equal radius regions are shown on the left and are overlaid on the X-ray surface brightness image. Figure \ref{fig:multi} shows that the X-ray surface brightness profile is steeper toward the NW (in the direction of the radio arc) than toward the SE. Moreover, we know that the Sunyaev-Zeldovich SZ emission is not coincident with the brightest cluster galaxy (BCG) position, and the BCG and the SZ peak are at a projected distance of $\sim$130\,kpc along the NW.
We should also note that the peak of the X-ray emission is closer to the BCG ($\sim$12\,kpc). 
In addition, we measured a metallicity of $Z\sim1.0$ at $\rm R < 75 kpc \simeq 0.08 R_{500}$, a concentration of $c\sim0.1$, and a centroid shift of $w=7.2\cdot10^{-3}$. These values are consistent with the typical values observed for relaxed clusters (e.g., \citealt{Mernier2018}, \citealt{LL19}, \citealt{Lovisari2017ApJ...846...51L}).

The conclusion about A3718 being a relaxed cluster would favor the second hypothesis made in the introduction: The radio arc could be a tail associated with S5RG. The shape of the arc could be due to the motion of the host galaxy S5 in the ICM.
Moreover, the width of the tail remains uniform along its full length, in contrast to other head-tail radio galaxies where the tail broadens, going away from the host due to adiabatic expansion \citep[see e.g., NGC\,7385 and PKS\,B0053-015 in][]{Sebastian2017AJ....154..169S}.
Based on these observations, we explored the X-ray image further to understand whether the ICM has a role in shaping this structure.

\subsection{X-ray spectral maps}
We attempted to better investigate the possible interplay between the thermal gas and the radio arc through X-ray spectral maps. Following the same method used in \cite{Brienza2022}, we extracted the spectra from small regions. We used the weighted Voronoi tessellation binning algorithm by \cite{Diehl2006}, which is a generalization of the \cite{Cappellari2003} Voronoi binning algorithm, to obtain the sizes of the spatial bins.

The resulting temperature, density, pressure, and entropy maps are shown in Figure \ref{fig:voronoi} with EMU radio contours.
All of the images highlight the X-ray asymmetry observed in the original \emph{XMM-Newton} image.
In particular, lower values of both temperature and entropy to the NW in the region of the arc with respect to those observed on the opposite side of the cluster center can be seen. Even if these images suggest a connection between the radio arc and the ICM, we would need deeper X-ray and radio observations to understand the physics at play in this particular and intriguing case.

\section{Radio spectral index analysis}
The radio source spectral index behavior is a key tool for investigating the physics of a radio source and identifying where particles are injected and how they age. In this section, we derive the spectral index properties of the extended radio source.

\subsection{ASKAP spectral index image}
We analyzed the resolved spectral index of the radio source considering the EMU and POSSUM images at 20\,arcsec. The spectral index map and uncertainty were computed as:
\begin{equation}
    \rm \alpha=-\frac{\log(B_1/B_2)}{\log(\nu_1/\nu_2)},
\end{equation}
\begin{equation}
    \rm \Delta\alpha=\frac{1}{\log(\nu_1/\nu_2)} 
    \sqrt{\left ( \frac{\Delta B_1}{B_1}  \right )^2 + \left ( \frac{\Delta B_2}{B_2}  \right )^2 },
\end{equation}
where $B_1$ and $B_2$ represent the surface brightness at frequency $\nu_1$ and $\nu_2$, respectively (i.e., the EMU and POSSUM frequencies). The results are shown in Figure \ref{fig:askap_spix} with the EMU contours overlaid.
The extended radio source shows a steepening spectral gradient along its length, from a spectral index $\alpha \simeq 0.6$ at the center near S5RG to a value of $\alpha \simeq 4$ at the furthest distance from S5RG (excluding regions corresponding to S1).

\begin{figure*}
    \centering
    \includegraphics[width=0.3\textwidth]{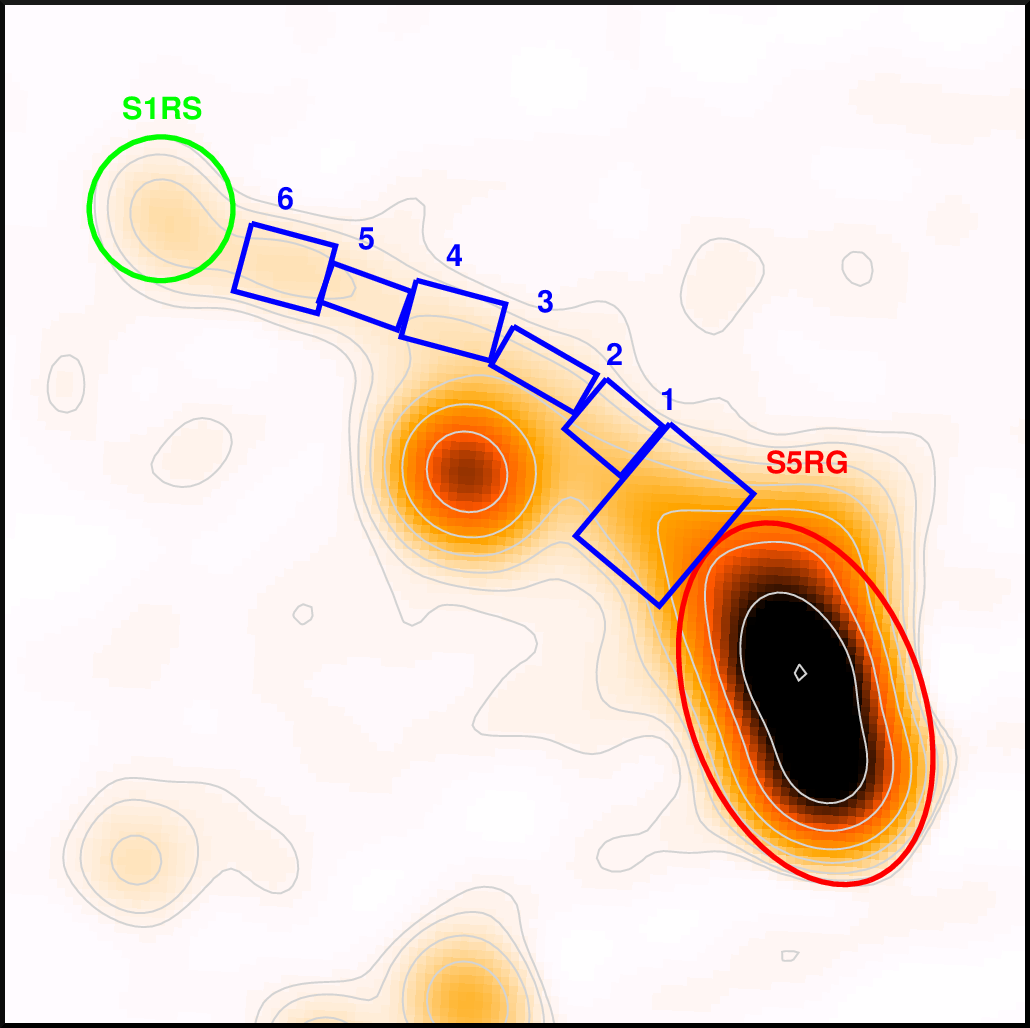}
    \includegraphics[width=0.43\textwidth]{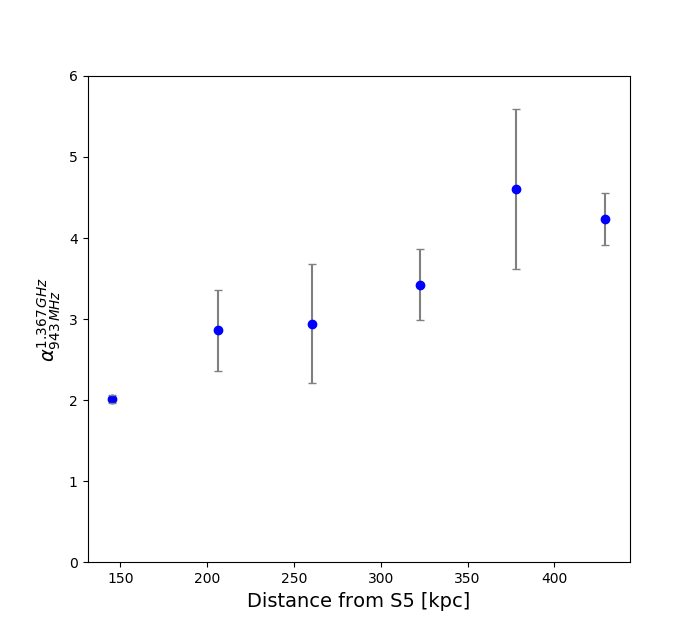}
    \caption{Spectral index gradient along the radio arc. Left: 943 MHz EMU image. Right: ASKAP spectral index as a function of the distance from the galaxy S5 measured in the six blue boxes shown in the top image.}
    \label{fig:arc}
\end{figure*}

The spectral steepening along the arc can be better appreciated by examining the profile in subregions. This is presented in Figure \ref{fig:arc}, where we derived the spectral index using the integrated flux density of our EMU and POSSUM maps. The profile shown in Fig. \ref{fig:arc} echoes the behavior seen in the maps presented in Figure \ref{fig:askap_spix}. Considering the spectral index trend along the arc and assuming that the cluster is relaxed, it is possible that the arc is indeed a tail associated with the S5 galaxy.

We note that the northern cluster galaxy, S1, appears to show a spectral index gradient toward the center, from a value of $\alpha \simeq 0.8$ to $\alpha \simeq 3$ and with a typical uncertainty $\Delta \alpha \lesssim 1$. Due to the steep-spectrum nature of the tail and the fact that these components are blurred in our EMU maps but not in our POSSUM maps, we suggest that this apparent gradient could be the result of blending between radio emission associated with S1 and the arc.

Finally, we also note that the BCG region has a steep spectral index of 1.5 at the peak of the surface brightness and flatter values toward northwest and southeast, where we measure a spectral index of about one. This spectral index behavior could be due to the blending of the central cluster sources.

\begin{figure*}
    \centering
     \includegraphics[width=0.49\textwidth]{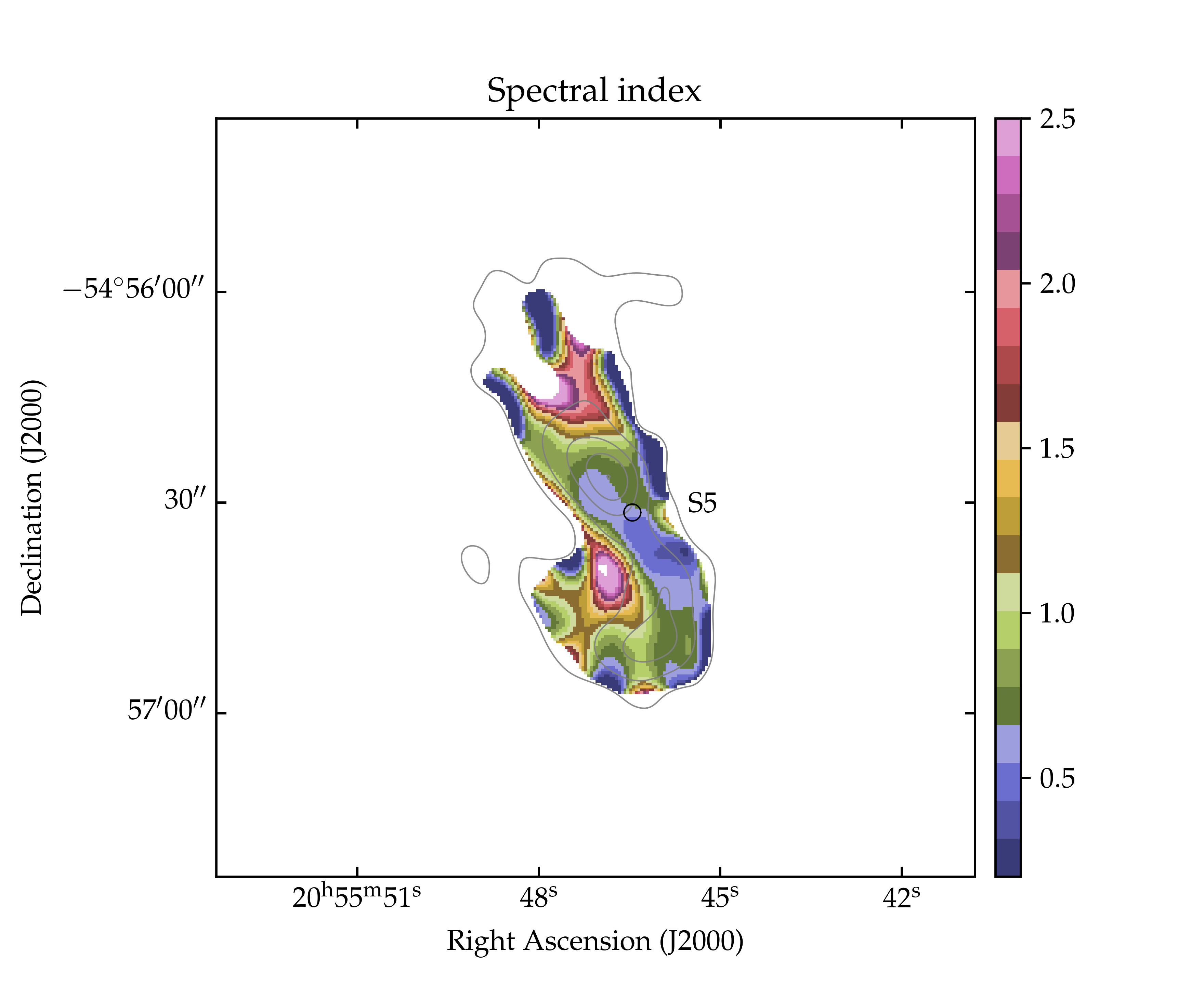}
     \includegraphics[width=0.49\textwidth]{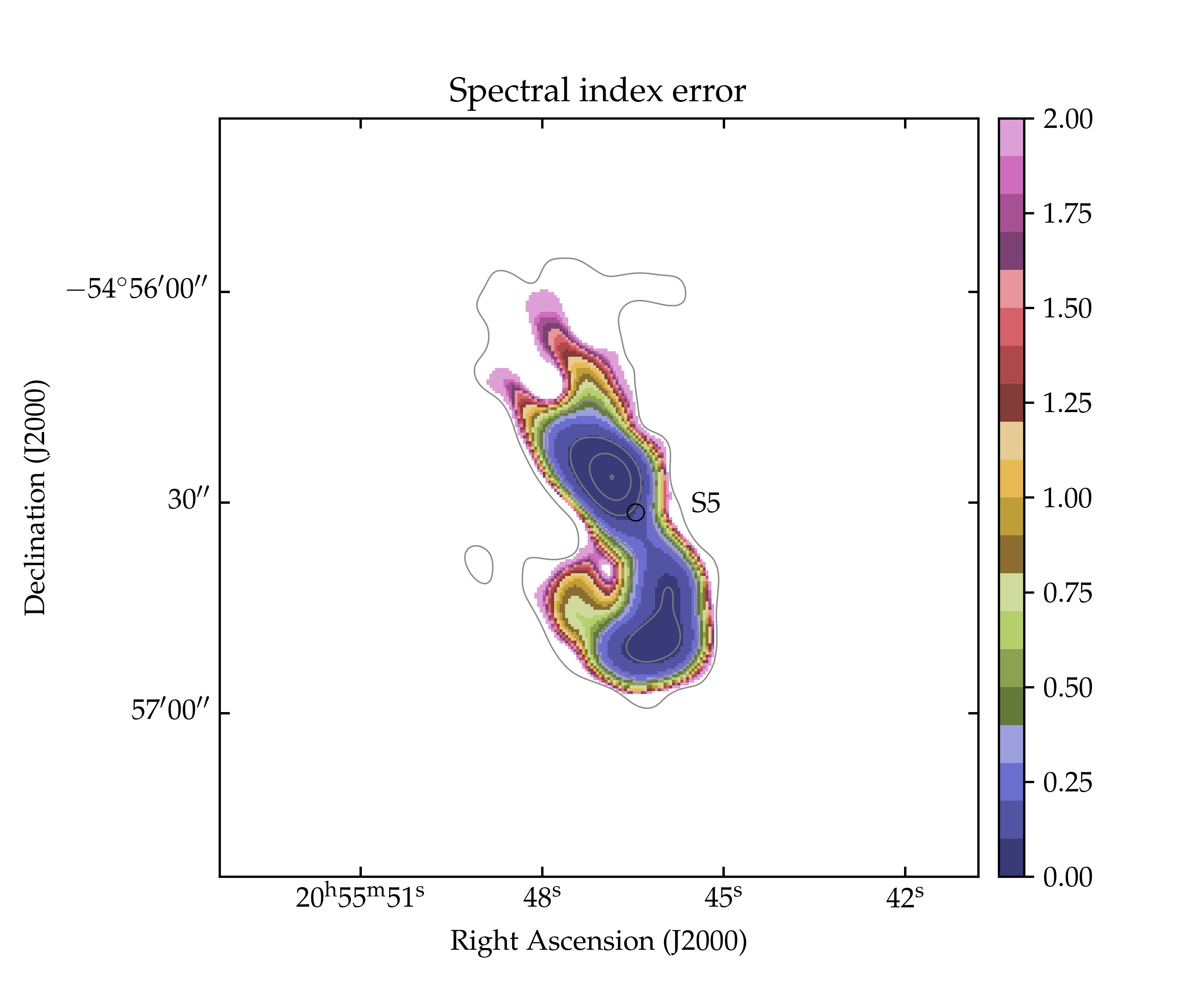}
    \caption{ATCA spectral index map and uncertainty map computed between 5.5 GHz and 9.0 GHz at 5\,arcsec of resolution.}
    \label{fig:atca_spix}
\end{figure*}

\subsection{ATCA spectral index image}
The interpretation proposed in the previous section seems consistent with the spectral index image made between 5.5 and 9.0\,GHz after convolving the ATCA images to a common resolution of 5\,arcsec with a circular beam. The results of this process are shown in Figure \ref{fig:atca_spix}.

The spectral index steepens as the distance from S5 increases both the north and the south. This strongly supports our hypothesis that S5 is the host of the radio emission S5RG. To the south, the spectral steepening follows the morphology of the radio emission, as the spectral index is increasingly steep along the presumed axis of the jet or lobe as it travels south before bending to the east and then northeast.

It is important to note that the western structure of the fork tail shows a flat spectral index of about 0.5. Although the associated error is quite large ($\sim$1.6), we can argue that some mechanism is reenergizing the particles in this region, maybe the same mechanism that is shaping the western structure.

\section{Summary and discussion}
In this paper, we conducted a multifrequency study of the galaxy cluster A3718. This far southern cluster, at a declination of $-54$\,degrees, has been poorly studied so far, and no previous radio frequency studies are present in the literature. Our interest was triggered by a radio arc located near the cluster center and detected for the first time during the EMU Phase I Pilot Survey. The radio arc, due to its morphology, could be either a radio relic or a tail associated with a radio galaxy.

Optical and X-ray data suggest that the cluster is in a relaxed dynamical state, which is consistent with the interpretation reported by \citet{zenteno2020}. These authors carried out a joint SZ, X-ray, optical analysis of the dynamical state of 288 massive galaxy clusters. In particular, they concluded that A3718 (which they referred to as SPT-CLJ2055-5456) is a relaxed cluster. Thus, the interpretation of the observed radio arc as being a relic is disfavored by their conclusion. Indeed, radio relics are usually found in merging clusters where shock waves can amplify magnetic fields and accelerate particles. However, the X-ray image shows a slight asymmetry with a flattened NW edge that could indicate an interaction with the complex radio source. Moreover, we observed a clear asymmetry in the X-ray spectral maps, suggesting that the cluster is not entirely relaxed. In particular, we note that the temperature and entropy maps show lower values NW from the X-ray peak in the region of the radio arc, while they increase in the opposite direction with respect to the X-ray peak. This asymmetry suggests a connection between the radio arc and the ICM. However, we do not have enough resolution and sensitivity to draw strong conclusions. Deeper observations with \emph{XMM-Newton} and/or \emph{Chandra} would be required.

By analyzing ASKAP data from the pilot  EMU and POSSUM surveys at 943\,MHz and 1.367\,GHz, respectively, and targeted ATCA observations at 5.5\,GHz and 9.0\,GHz, we have discovered that the radio emission observed in the EMU image is composed of a compact radio source associated with the optical galaxy S1, an arc of radio emission having a length of $\sim$370\,kpc and a width of $\sim$25\,kpc, and a southern radio galaxy (S5RG) with two lobes bending in the arc direction and extending for $\sim$200\,kpc. The radio arc is extremely thin. Therefore, it is possible that we are detecting only the brightest patches associated with this source.

The ASKAP and ATCA images revealed two additional intriguing structures. The first structure can be seen in the ASKAP image northeast from S5RG and seems to be a jet that either failed to expand or faded away due to energy losses. The second structure is a forked tail observed in our ATCA images toward the northwest of the northern lobe of S5RG.

The ATCA polarized intensity image shows B-vectors aligned with the direction of the lobe or jet, and this supports the picture proposed. Curiously, the western part of the fork tail shows a transverse B-field. In the RM image, we predominantly found positive/negative values associated with the southern/northern lobes of S5RG. This suggests that there is a different portion of magneto-ionic plasma in front of the lobes. One potential cause of the RM difference between the lobes could be their different distance along the line of sight, and thus the observed emission passes along different path lengths through the ICM. This would then imply that the radio galaxy and the arc extend for greater distances than implied by the projected 2D size.

Considering the ASKAP and ATCA spatially resolved spectral index images, the radio arc seems to be a tail associated with S5RG. Indeed, we observe a steepening of the spectral index from the S5RG, where we measure $\alpha_{943 {\rm MHz}}^{1.367 {\rm GHz}} \approx 0.6$ along the arc. Toward the end of the arc, the spectral index reaches very steep values with $\alpha_{943 {\rm MHz}}^{1.367 {\rm GHz}} \approx 4$.

From these results we have computed a first order estimate of the radiative age of the arc, assuming that particles age following a JP model \citep{jaffe1973}. Considering a magnetic field strength that minimizes the energy losses (i.e., B=B$\rm_{CMB}/\sqrt{3}=2.4\,\mu$G) and a spectral injection index of 0.6, as measured in the region closest to the S5 galaxy in the EMU-POSSUM spectral index image, which is likely close to particle injection, we have simulated the evolution of the JP spectrum in time. According to our simulation, to observe a steepening of the spectrum up to the observed index $\alpha_{943 {\rm MHz}}^{1.367 {\rm GHz}} = 4$, the source age must be around 180~Myrs.

We then computed the path covered by the galaxy S5 through the ICM in this amount of time and compared it with the radio source observed length.
We do not know the velocity of the galaxy, but we can assume the equipartition condition between the ICM and the galaxy energy density \citep[see, e.g.,][]{boschin2018}:
\begin{equation}
    \sigma_{\rm v}^2 = (kT_{\rm X})/(\mu m_{\rm p}),
\end{equation}
where $\mu=0.58$ is the mean molecular weight and $m_{\rm p}$ the proton mass. Given the ICM temperature of $T_{\rm X}\sim4.4$\,keV, the resulting galaxy velocity dispersion is $\sigma_{\rm v} \sim 800$\,km/s.
Therefore, taking this value for the velocity of the galaxy, during 180\,Myrs it moved through the ICM to a distance of $l=v\cdot t$\,=\,140\,kpc, a factor of $\sim$3.3 smaller than the length of the radio arc plus the northern lobe of S5RG.

Indeed, to cover the observed distance, the galaxy S5 would need a time interval of about 600\,Myrs.
Such an age would imply a spectral index much steeper than the observed profile. Moreover, after this amount of time, we should not be able to detect the arc, due to its very low surface brightness. Crucially, however, we have noted that the arc is very narrow. Normally, tails associated with radio galaxies become broader due to the spontaneous adiabatic expansion of the plasma, unless some external interaction causes a recollimation \cite[see for example][]{Velovic2022MNRAS.516.1865V}.
The physical mechanism responsible for the characteristics of the observed radio arc is not clear, but we can examine three options: 1) The radio galaxy is ejecting magnetic field and particles at such high velocity that they can cover the observed arc length. 2) Instabilities between the tail and the ICM cause turbulence that may disrupt and re-accelerate dormant tail electrons. 3) The nonthermal plasma of the arc is being re-accelerated through weak shock(s) or compression.

Jet velocities can be extremely high in radio galaxies. \cite{Laing2002} estimated a jet velocity for 3C31 of the order of 10$^5$km/s nearby the AGN (tens of kpc).
\cite{oneill2019} produced a magneto-hydro-dynamical simulation of a narrow-angle tail galaxy assuming jet velocity of the order of $10^4$\,km/s and an ICM with similar density but with a pressure higher by a factor of ten with respect to A3718. After 540\,Myrs, the tails reach a length comparable to our radio source, with a steepening of the spectral index from the core along the tails. 
However, the spectral index steepens to a maximum of $\alpha_{950 {\rm MHz}}^{1.4 {\rm GHz}} = 2$, while in our case the spectral index at the end of the arc is significantly steeper, as we observed values up to $\alpha_{943 {\rm MHz}}^{1.367 {\rm GHz}}=4$. A combination of the galaxy motion through the cluster and a jet velocity could be enough to explain the presence of the arc and its shape, although only dedicated numerical simulations could give us a definitive answer. Indeed, if this is the case, we would expect to see a combination of freshly injected particles and aged ones at greater distances from the hosting galaxy. This would probably result in a spectral index flatter than what we observed, as also shown in the aforementioned simulations.

The second option for explaining the physical mechanism has been proposed in several cases \citep{dega2017,Wilber2018,muller2021,Botteon2021,Ignesti2022}. 
Turbulence is indeed expected in tails of radio galaxies due to the interaction with the ICM.
This mechanism can re-accelerate particles and prevent their aging, causing a flattening of the spectral index, especially at greater distances from the host, where radiative losses are important. This explanation has been tentatively reported by \cite{Sebastian2017AJ....154..169S}. However, even this scenario is disfavored by the observed spectral index gradient, at least at this resolution that corresponds to about 33\,kpc.

The third scenario also implies a flattening of the radio spectrum. Weak shocks can reenergize high-frequency particles, while compression results in a shift of the spectrum, and both mechanisms can leave traces in the X-ray images. According to the X-ray spectral maps, we observed only a possible enhancement of the pressure, and this scenario also implies an increase in the temperature and entropy. Deeper X-ray data as well as higher-resolution radio observations at low frequencies would be required to probe this scenario further.

\section{Conclusions}
In this work we conducted a multifrequency study of the galaxy cluster A3718 and, in particular, of a complex radio source detected for the first time during the EMU phase I survey.
The radio source appears as an extremely thin ($\sim$30\,kpc) emission very close to the cluster center ($\sim$12\,kpc) and extending for $\sim$612\,kpc. The morphology is similar to previously detected radio relic structures. However, the optical and X-ray data suggest that A3718 shows an overall relaxed morphology, which is in conflict with the general idea that radio relics trace shock waves propagating in merging clusters.

By analyzing the 943\,MHz and 1.367\,GHz ASKAP and the 5.5\,GHz and 9.0\,GHz ATCA data, both in total intensity and polarization, we concluded that the extended radio source is composed of a compact radio source in the north, a thin radio arc, and a southern radio galaxy (S5RG). The radio arc and S5RG cover a projected distance of $\sim$370\,kpc and $\sim$200\,kpc, respectively.
The radio arc is likely a tail associated with S5RG, although we did not detect a clear connection between these two structures at high resolution in our ATCA images. This is not surprising since the arc has a very steep spectral index, with $\langle \alpha_{943 {\rm MHz}}^{1.367 {\rm GHz}} \rangle = 2-4,$ and it is very faint (S$\rm_{943\,MHz}=73\,\muup Jy/arcsec^2$).

Considering the arc and the radio galaxy S5RG as a unique system, we concluded that the motion of the galaxy through the ICM is not enough to produce the observed radio length. Therefore, either the radio jet is ejected with a velocity of the order of 10$^4$km/s or the arc is receiving an energy boost from the ICM. With the available data, we cannot tell whether the arc is powered by turbulence induced by instabilities between the arc and the ICM, compression, or weak shock(s). Deeper X-ray and radio observation would allow us to finally shed light on the mechanism responsible for this ultra-steep and very thin radio arc near the center of A3718.

\begin{acknowledgements}
We thank the anonymous reviewer for the useful comments and suggestions. 
FL acknowledges financial support from the Italian Minister for Research and Education (MIUR), project FARE, project code R16PR59747, project name FORNAX-B. 
FL acknowledges financial support from the Italian Ministry of University and Research $-$ Project Proposal CIR01$\_$00010. 
MB acknowledges financial support from the program UNESCO-L'Oréal Italia per le Donne nella scienza, from the ERC-Stg ``DRANOEL", no. 714245, from the ERC-Stg ``MAGCOW", no. 714196, from the agreement ASI-INAF n. 2017-14-H.O and from the PRIN MIUR 2017PH3WAT “Blackout”.
AB and CJR acknowledge support from the ERC Starting Grant `DRANOEL', number 714245.
AB and CS acknowledges financial support from the Italian Minister for Research and Education (MIUR), project SMS.
The Australian SKA Pathfinder is part of the Australia Telescope National Facility (https://ror.org/05qajvd42) which is managed by CSIRO. Operation of ASKAP is funded by the Australian Government with support from the National Collaborative Research Infrastructure Strategy. ASKAP uses the resources of the Pawsey Supercomputing Centre. Establishment of ASKAP, the Murchison Radio-astronomy Observatory and the Pawsey Supercomputing Centre are initiatives of the Australian Government, with support from the Government of Western Australia and the Science and Industry Endowment Fund. We acknowledge the Wajarri Yamatji people as the traditional owners of the Observatory site. The POSSUM project (https://askap.org/possum) has been made possible through funding from the Australian Research Council, the Natural Sciences and Engineering Research Council of Canada, the Canada Research Chairs Program, and the Canada Foundation for Innovation.
The Australia Telescope Compact Array is part of the Australia Telescope National Facility (https://ror.org/05qajvd42) which is funded by the Australian Government for operation as a National Facility managed by CSIRO.
We acknowledge the Gomeroi people as the traditional owners of the Observatory site.
This research has made use of the CIRADA cutout service at URL cutouts.cirada.ca, operated by the Canadian Initiative for Radio Astronomy Data Analysis (CIRADA). CIRADA is funded by a grant from the Canada Foundation for Innovation 2017 Innovation Fund (Project 35999), as well as by the Provinces of Ontario, British Columbia, Alberta, Manitoba and Quebec, in collaboration with the National Research Council of Canada, the US National Radio Astronomy Observatory and Australia’s Commonwealth Scientific and Industrial Research Organisation.
\end{acknowledgements}

%
%
\bibliographystyle{aa}
\bibliography{corr}
\end{document}